\documentclass{article}

\usepackage{aimc2025}

\usepackage[utf8]{inputenc} % allow utf-8 input
\usepackage[T1]{fontenc}    % use 8-bit T1 fonts
\usepackage{hyperref}       % hyperlinks
\hypersetup{
  colorlinks   = true, %Colours links instead of ugly boxes
  urlcolor     = blue, %Colour for external hyperlinks
  linkcolor    = black, %Colour of internal links
  citecolor   = black %Colour of citations
}
\usepackage{url}            % simple URL typesetting
\usepackage{booktabs}       % professional-quality tables
\usepackage{amsfonts}       % blackboard math symbols
\usepackage{nicefrac}       % compact symbols for 1/2, etc.
\usepackage{microtype}      % microtypography
\usepackage{graphicx}       % figures
\usepackage{amssymb,amsmath,epsfig}
\usepackage{booktabs}
\usepackage{makecell}
\usepackage{diagbox}
\usepackage{multirow}
\usepackage{amsthm}

\usepackage{caption}
\usepackage{subcaption}

\title{Learning Relationships between \\ Separate Audio Tracks for Creative Applications}

\author{%
  Balthazar Bujard\\
  ISMM Team\\
  STMS Lab Ircam - CNRS - Sorbonne Université\\
  \And
  Jérôme Nika \\
  ISMM Team\\
  STMS Lab Ircam - CNRS - Sorbonne Université\\
  \And
  Fédéric Bevilacqua \\
  ISMM Team\\
  STMS Lab Ircam - CNRS - Sorbonne Université\\
  \And 
  Nicolas Obin \\
  Sound Analysis and Synthesis  Team\\
  STMS Lab Ircam - CNRS - Sorbonne Université\\
  \\
  1 Place Igor Stravinsky, 75004 Paris, France \\
  \texttt{firstname.lastname@ircam.fr} \\
}

\begin{document}

\maketitle

\begin{abstract}
  This paper presents the first step in a research project situated within the field of musical agents. The objective is to achieve, through training, the tuning of the desired musical relationship between a live musical input and a real-time generated musical output, through the curation of a database of separated tracks. We propose an architecture integrating a symbolic decision module capable of learning and exploiting musical relationships from such musical corpus. We detail an offline implementation of this architecture employing Transformers as the decision module, associated with a perception module based on Wav2Vec 2.0, and concatenative synthesis as audio renderer. We present a quantitative evaluation of the decision module’s ability to reproduce learned relationships extracted during training. We demonstrate that our decision module can predict a coherent track B when conditioned by its corresponding "guide" track A, based on a corpus of paired tracks (A, B).  %We highlight the difficulty of learning musical relationships and formulate the guidelines necessary for future works.
\end{abstract}

\section{Introduction}\label{sec:introduction}

\subsection{Problematic: tuning the relationship between inputs and outputs in Musical Agents}\label{sec:context}
%\note[BB]{changer le titre pour parler de problématique. introduire ce qui est interessant c'est les relations entre entrée etr sortie}

The present work is placed in the context of a broader project within the field of Musical Agents (MA), defined by \citet{musical_agents} as  "[agents] that partially or completely automate musical creative tasks". Our long term objective is to achieve, through training, the desired tuning between a live musical input — captured via a machine listening module — and a real-time generated musical output, through the curation of a database of separated tracks. Its applications and areas of investigation are considered in both compositional and improvisational contexts. The architecture we consider is composed of: an \emph{action} module responsible for musical synthesis guided by a \emph{perception} module implementing machine listening techniques to extract information from a performer’s playing on stage, and a \emph{decision} module creating a mapping from perception to action.

Our approach is closely related to and inspired by earlier performance-driven MA(s). ImproteK \citep{improtek} is a reactive MA designed for live improvisation. Somax \citep{somax} introduces long-term planning capabilities through symbolic scenarios. MASOM \citep{masom} and SpireMuse \citep{spiremuse} focus on live improvisation, incorporating affective dimensions into their machine listening. tiNNbre \citep{tinnbre} integrates neural networks into its decision module. Lastly, Construction III offers a performance-driven environment that leverages transformative and sequenced techniques \citep{cypher}. All these agents implement their action module through corpus-based concatenative synthesis.\footnote{With the exception of tiNNbre v2, which generates audio through a Griffin-Lim re-synthesis algorithm \citep{griffin-lim}.} This technique employs a musical corpus stored in memory, composed of pairs of (segment, label), i.e. (content, symbol), to generate audio.

Particularly, Dicy2 \citep{dicy2}, our action module (see Section \ref{sec:action_implementation}), is a \textit{scenario}-based MA combining the reactive-listening of Somax and the long-term scenario planning of ImproteK. Dicy2 utilizes concurrent dynamic calls to symbolic scenarios, i.e. sequence of symbols, to generate audio with corpus-based concatenative synthesis. The scenario can be generated from the audio analysis of a musician's play for real-time improvisation, or specified during offline music composition. This system learns temporal relationships within its corpus and, when providing a scenario, it generates the optimal sequence of content. 

%\subsection{Motivation: musical responses beyond matching}

During generation, the decision module selects the segments in the corpus to be played. In general, this selection is primarily based on a combination of probabilistic models and pattern matching, with the objective of creating an optimal direct mapping between perception and action. 

%\note[BB]{Le cas de MASOM peut poser problème. Il fait appel a des dimensions "affectives", pleasantness et eventfullness. Ce sont effectiveemnt des dimensions haut niveau, mais elles sont composées de combinaison linéaires de paramètres bas-niveau, déterminées par regression linéaire.}

Usually, the perception module implements real-time audio stream analysis based on audio descriptors, such as picth, spectral centroid, MFCCs. %One could argue that MASOM implements some high-level dimensions ("pleasantness" and "eventfulness") to its listening module, still these dimensions are defined as a combination of low-level audio descriptors \citep{masom}. 
While classic audio descriptors are interpretable, they are limited in their ability to encode high-level musical information \citep{Audio_representations_for_DL}. %\note[BB]{Du coup cette conclusion peut etre débatue, puisque a partir de combinaisons de params acoustiques simples on peut définir des dimensions haut niveau. Cependant MASOM construit ces paramètres avec un petit jeu de données labelisées par des humains (20 humains et 125*6[s] de données), et donc leur pertinence (?) peut être discutée}. 

%This paper focuses on the decision module of MAs, which generally makes use of matching rules between the generated signal and input signal. This entails the construction of musical responses based on principle of similarity. \note[BB]{Therefore, integrating a decision module capable of learning relationships from a set of examples is of great interest in this field; expanding the creative processes from such systems.}

To summarize, MAs generally make use of matching rules between the generated signal and input signal. This implies the specification by hand of the musical behaviour of the agent, i.e., the explicit definition of the audio description and pre-defined scenarios. Though this offers a lot of possibilities, this could be a limitation for more complex relationships. This paper addresses the possibility to automatically learn musical relationships from a set of examples, substituting pre-defined specifications with behaviours directly learned from examples. This is done by exploring the capacities of state-of-the-art neural networks architectures both in the field of audio and music processing, as described in the following section. Ultimately, we aim at expanding the customizability and the versatility of existing MAs.

%\note[JN]{OU BIEN : To summarize, MAs which generally makes use of matching rules between the generated signal and input signal. This entails the construction of musical responses based on principle of similarity. This paper adresses therefore the issue of integrating a decision module capable of learning relationships from a set of examples to expand the customizability of such systems.}

%\note[BB]{Les agents musicaux, quoique établis dans la littérature et sur la scène artistique, ... en gros lea agents musicaux ne peuvent aps être accordés. le principe de matching bien établi est simple et efficace, mais il limite la capacité des systèmes a s'adapter à un artiste, style etc. Ca veut pas dire que ces agents ne peuvent pas etre allignés avec les volontés artistiques/musicales de l'utilisateur, mais il serait pertinent de chercher un moyen d'accorder ces agents a un style de jeu donné. cela permettrait d'étendre l'expressivité de ces agents.}

%\note[BB]{Furthermore, the set of acoustic descriptors directly influences the matching process, limiting the construction of responses following more complex musical dimensions. For some agents, e.g. Dicy2, the choice of descriptors is a user-defined parameter so the user must know which acoustic feature is of interest for its application. Removing this parameter alleviates the mental load of the artist but restricts its customization. Providing a description module incorporating high-level musical information could have some benefits.}

%Our motivation is therefore to extend both the perception and decision engines.

\subsection{Our Approach: tuning Musical Agents' responses through examples}

Our motivation is  to extend both the perception and decision modules. To improve Musical Agents' response mechanism we decide to, first, implement the perception module with neural audio encoders to represent the audio signals with high-level musical information, and secondly, integrating a decision module capable of formulating responses based on learned relationships between separate musical tracks on a symbolic level. Additionally, we propose a novel task for the decision module focused on learning correspondences between inputs and outputs through example-based training. 

%First, the perception leverages advances in neural audio encoders to represent the audio signals with high-level musical information. Second, the response mechanisms should function beyond a simple matching paradigm. Particularly we aim at integrating a decision engine capable of formulating responses based on learned relationships between separate musical tracks on a symbolic level. For this, we propose a novel task for the decision engine, focused on learning correspondences between inputs and outputs through example-based training. 

Specifically, we propose to evaluate the ability of such decision module to learn and generate coherent symbolic relationships based on a corpus of paired tracks (A, B) using what we call a \emph{re-generation} task. The goal is not to assess expressive or creative output \textit{per se}, but to quantitatively measure the model’s capacity to learn and generate similar structural relationship between symbolic representations of separate tracks. While creative generation ultimately requires subjective evaluation, this re-generation task provides an initial, objective assessment of the decision's capacity to learn and generalize musical relationships.

This article presents the general framework of the project, the proposed methodology, and an initial quantitative evaluation based on this re-generation task.

The present work proposes the following contributions to the field of Musical Metacreation, Musical Agents, and AI Creativity:
\begin{itemize} 
    \item The formalization of a high-level architecture for relationship-driven MA(s). 
    \item An offline implementation of this architecture using Deep Learning for both perception and decision-making components. 
    \item The design of a quantitative evaluation framework based on a re-generation task to assess the decision module's ability to internalize musical relationships. 
\end{itemize}

In the following, applications of Deep Learning in music are presented in Section \ref{sec:related_work}. Then the high-level architecture of the proposed Musical Agent is formalized in Section \ref{sec:architecture}. The proposed implementation of this architecture is detailed in section \ref{sec:model}. Section \ref{sec:method} presents the methodology applied during this research: dataset curation, training procedure, evaluation protocol, and other relevant details. The results are presented in Section \ref{sec:results} and discussed in section \ref{sec:discussion}. Finally, key findings and future work are presented in section \ref{sec:conclusion}.

%In the following, related work is presented in the section \ref{sec:related_work}, and research on interactive music systems (Section \ref{sec:music_interactive_systems}) and artificial intelligence applied to music (Section \ref{sec:gen_ai_in_music}) is presented. The former serves primarily to present the main motivations for our work, and the latter to highlight the distinctive features of our system. The high-level/abstract architecture of the proposed framework is then presented in Section \ref{sec:architecture}. The specific implementation of the audio-audio coupling system is detailed in section \ref{sec:model}. Section \ref{sec:method} presents the methodology applied during this research; from dataset curation, evaluation protocol, through training procedure, and all details covering our methodology. The results are presented in Section \ref{sec:results} and discussed in section \ref{sec:discussion}. Finally, key findings and future work are presented in section \ref{sec:conclusion}.

\section{Related Work}\label{sec:related_work}

%This work is firmly embedded in the interdisciplinary field of deep learning (DL), Musical Metacreations (MuMe), and specifically Improvising MA(s). The previous section presented a description of Improvising MA(s), providing some relevant examples of such agents, mainly as a way to expose our key motivation driving our work; among other considerations relevant to the proposed framework and evaluation protocol. Since our system implementation is mostly built upon DL architectures, a more extensive overview of the state of the art of DL in the musical field is presented. The ensuing discussion will focus on two key areas: generative models and neural audio encoders. One to highlight the distinctive features of our system and the latter to provide a formal overview of the technology employed in our MA.

This work lies at the intersection of Deep Learning (DL), Musical Metacreation, and more specifically, Musical Agents. Given that our approach is primarily built upon DL architectures, we begin by presenting a relevant overview of the state of the art in DL within the musical domain. The discussion is organized around two core areas: neural audio encoders, which provide a formal foundation for the technologies underpinning our perception module implementation, and generative models, which contextualize the distinctive features of our approach.

\subsection{Neural audio encoders for musical Signals}\label{sec:music_encoders}

%Therefore, the focus is directed towards neural audio encoders that effectively represent meaningful musical information.

%Neural audio encoders' goal is to project a stream of audio into a compressed and generalized sequence of vectors representing relevant audio information \citep{audio_self_supervised}. In the following, we present a few interesting neural encoders adapted for musical signals. MusicBERT \citep{musicbert} is a bidirectional Transformer-Encoder architecture pre-trained congruently on an alignment and reconstruction tasks to learn musical structure and co-relations between different music. EnCodec \citep{encodec} can also be used as a powerful music encoder, it represents audio information in a hierarchical and quantized stream of vectors that has been used to represent music \citep{audiolm}. Wav2Vec 2.0 \citep{wav2vec_speech} is a bidirectional Transformer-Encoder that constructs contextualized representations with a contrastive Masked Language Modeling pre-training objective. It has successfully been adapted to music data \citep{wav2vec_music} and its representations have been evaluated on pitch and instrument classification, obtaining competitive result against task-specific models.

Neural audio encoders aim to project audio streams into compressed vector sequences that capture relevant musical information \citep{audio_self_supervised}. Below, we highlight key encoders adapted for musical signals. MusicBERT \citep{musicbert} is a bidirectional Transformer encoder pre-trained on alignment and reconstruction tasks to learn musical structure and inter-relations across pieces. EnCodec \citep{encodec}, though designed for signal compression, also serves as a powerful encoder by producing a hierarchical, quantized vector stream which has been employed in music modeling \citep{audiolm}. Wav2Vec 2.0 \citep{wav2vec_speech}, a contrastively trained Transformer encoder for contextual audio representation, has been successfully adapted to music \citep{wav2vec_music}, achieving competitive results in pitch and instrument classification compared to task-specific models, attesting for its generalization capabilities.

We propose to take advantage of such encoders to enhance the role of machine listening in the creative process. Rather than encoding the audio signal with raw audio descriptors, our perception module would encode high-level musical information so that the decision module can learn high-level musical relationships.

%\note[BB]{our approach takes advantage of such encoders to improve the perception module. Rather than encoding the audio signal with audio descriptors, this approach encodes high level musical information, which enables the decision module to learn high-level musical relationships.}

%\note[JN]{OU BIEN : We propose to take advantage of such encoders to enhance the role of machine listening in the creative process. Rather than encoding the audio signal with raw audio descriptors, our perception module would encode high-level musical information so that the decision module can learn high-level musical relationships.}

%small conclusion and opening for architecture 
%We outlined the key components of IMS that motivate the integration of a decision stage into its architecture, enabling the formulation of musical responses based on complementarity. To contextualize our approach, we reviewed relevant research on music generation using deep learning, emphasizing what sets our work apart. Additionally, we introduced various music encoders that could contribute to our research. The following section provides a detailed overview of our framework's high-level architecture.

\subsection{Generative AI for music generation}\label{sec:gen_ai_in_music}

% we want to present accompaniment generation systems, symbolic sequence modeling and audio encoding. particularly present diff-a-riff, SingSong (and AudioLM) and Transformers for symbolic sequence generation
%To the best of our knowledge, this framework has never been explored before. The particularity of our system is that it operates on a symbolic level to learn musical relationships between \textit{related} tracks and generate responses based on complementarity. 
Our work is related to research on accompaniment generation systems conditioned by an audio input, namely \textit{guiding input} or \textit{musical context}, and systems exploiting Language Modeling  approaches for music generation, i.e. using symbols/tokens to represent audio.

The field of neural audio codecs is thriving \citep{soundstream, encodec, dac, soundstorm, snac}, with decoders enabling high-quality music generation from token streams (e.g. RAVE \citep{rave}, R-VAE \citep{r_vae}, SingSong \citep{singsong}, Jukebox \citep{jukebox} and Jukedrummer \citep{jukedrummer}. Notably, SingSong generates instrumental tracks conditioned on vocal input, adapting AudioLM \citep{audiolm} to model instrumental tokens from vocal tokens — a methodology closely related to ours, where the symbolic sequence of track A is modeled from that of track B. Transformers \citep{attention} have shown strong performance in symbolic music — score-based or MIDI — generation \citep{museformer, popmag, musictransformer}, and several works successfully combine them with audio codecs to generate audio \citep{musicgen, musicgen_stem, coco_mulla}. Diff-A-Riff \citep{diff_a_riff} offers conditional accompaniment generation via a dual conditioning scheme: musical context and/or direct control through textual prompts or audio references, enabling fine-grained guidance over the generated accompaniment.

The architecture we propose below is similarly conditioned by an audio input. Its distinctive feature is that it focuses not on generating audio, but on producing a high-level symbolic specification for audio generation. 
It uses deep learning to learn and exploit relationships between symbolic sequences extracted from curated paired audio files to generate spectifications to send to a corpus-based synthesis engine that will produce the actual musical response.

%\note[BB]{Our system is also conditioned by an input audio but does not synthesize audio. it uses deep learning to learn relationships between symbolic sequences extracted from audio streams with neural encoders and uses the corpus-based synthesis engine of musical agents to generate the musical response.}

%\note[JN]{OU BIEN : The architecture we propose below is similarly conditioned by an audio input. Its distinctive feature is that it focuses not on generating audio, but on producing a high-level symbolic specification for audio generation. 
It uses deep learning to learn and exploit relationships between symbolic sequences extracted from curated paired audio files to generate spectifications to send to a corpus-based synthesis engine that will produce the actual musical response.

\section{Architecture}\label{sec:architecture}

In the following Section, we propose a high-level architecture of a relationship-based Musical Agent and its mathematical formulation. This architecture is composed of three key components. A \textbf{Perception} module that converts a continuous audio signal into a sub-sampled discrete, i.e. symbolic, sequence. A \textbf{Decision} module that learns and exploits symbolic relationships between symbolic sequences. An \textbf{Action} module that converts a symbolic sequence back into audio.

\begin{figure}[ht]
    \centering
    \includegraphics[width=0.5\textwidth]{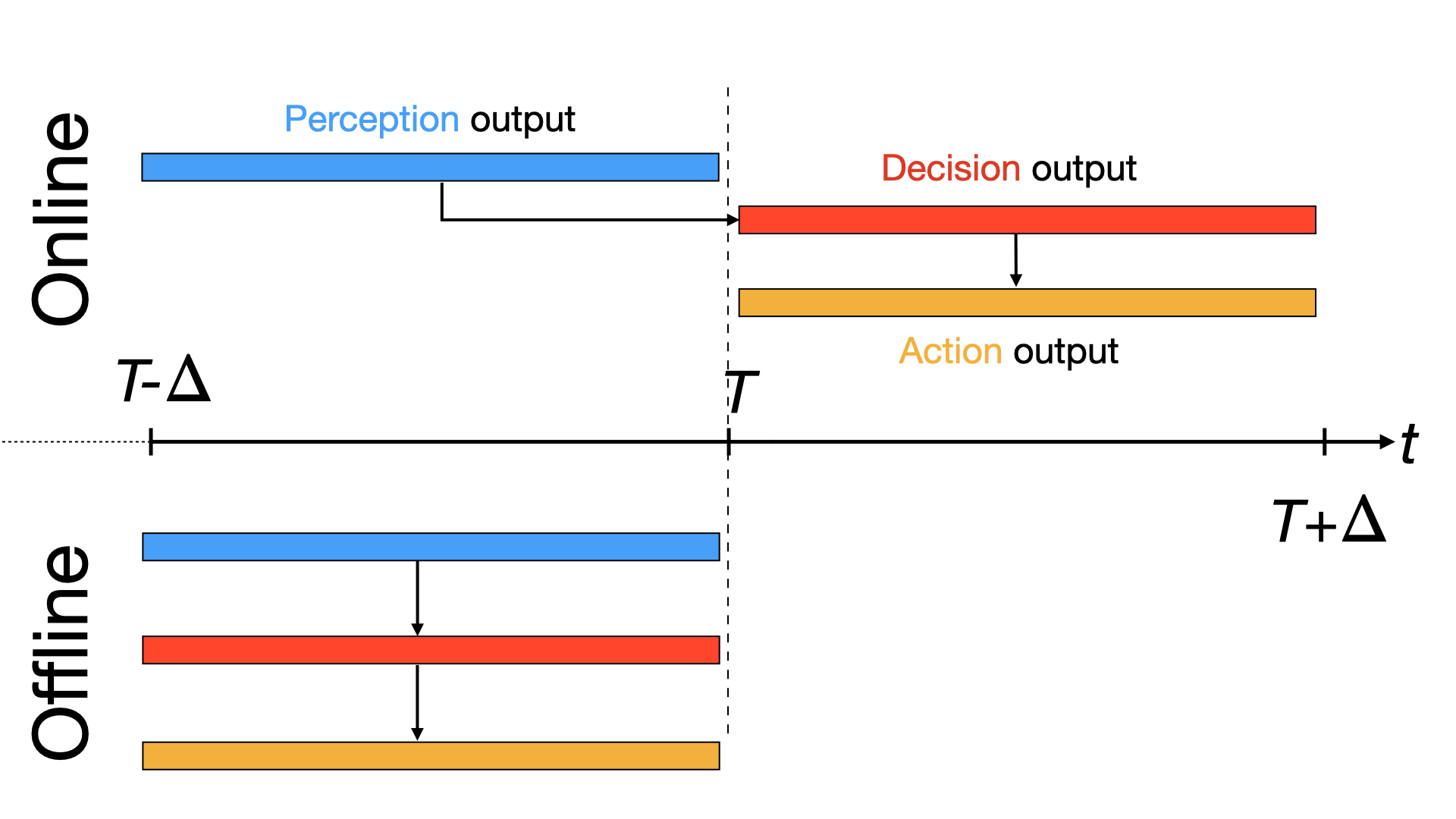}
    \caption{Online vs. offline framework for relationship-based Musical Agents. %The online framework requires to generate a symbolic specification ahead of the performance time T, while the offline framework processes all signals on the same time frame.
    }
    \label{fig:online_offline}
\end{figure}

Figure \ref{fig:online_offline} depicts the difference between an offline and online framework for Musical Agents. A finalized Musical Agent for real-time performance will involve a perception module based on machine listening to extract information from an audio stream over an interval [T-$\Delta$;T] in order to provide a symbolic representation that will be used by the decision module to generate a symbolic specification of the music to be generated over an interval [T;T+$\Delta$]. In this paper, we focus on the decision module and consider an offline architecture where the audio control stream is an audio file, and where the symbolic representation provided by the perception module over an interval [T-$\Delta$;T] serves as input to the decision module to generate a symbolic specification of the audio to be played over the same interval [T-$\Delta$;T].

The general offline framework consists in specifying the musical response, noted $\vec{\textbf{y}} = [y_1,y_2,...,y_{T_y}]$, corresponding to a guiding input $\vec{\textbf{x}}=[x_1,x_2,...,x_{T_x}]$, with $T_y \neq T_x$ in general. So the function $f$ realized by the agent is defined as: $y_1,y_2,...,y_{T_y} = f(x_1,x_2,...,x_{T_x})$. This function is modeled by the conditional probability $p(\vec{\textbf{y}}|\vec{\textbf{x}})$ defined in Equation \ref{eq:coupling}. The response $\vec{\textbf{y}}$ is generated auto-regressively by predicting the probability of the t-th sample ($y_t$) conditioned by previous samples ($\vec{\textbf{y}}_{<t}$) and the guiding input ($\vec{\textbf{x}}$). It is noteworthy that this probability is computationally intractable since it requires access to all previous samples of $\vec{\textbf{y}}_{<t}$ and $\vec{\textbf{x}}$, further validating the choice of modeling musical relationships on a subsampled, symbolic scale.

\begin{equation}\label{eq:coupling}
p(\vec{\textbf{y}}|\vec{\textbf{x}}) = \prod^{T_y}_{t=1}p(y_t|\vec{\textbf{y}}_{<t},\vec{\textbf{x}})
\end{equation}

\subsection{Perception: converting audio to a sequence of symbols }\label{sec:architecture_perception}
%Perception is a multi-sensory faculty, but in the present context we choose to limit it to listening. The first link of the coupling system is therefore an auditory \textit{Perception} module. 
The Perception module should capture complex musical information, beyond a simple choice of audio descriptors. Musical relationships, as they are approached in this work, are modeled through a symbolic alphabet. For this matter, the Perception module also converts the audio stream into a sequence of symbols. This layer of symbolic abstraction is achieved by means of a discrete musical alphabet. Obviously, the purpose of this alphabet should correspond to high-level classes of musical equivalences in order to allow the Decision module to extract relations between symbolic sequences.

Equation \ref{eq:formule_perception} describes the mathematical function realized by the Perception module : an audio sequence $\vec{\textbf{x}}$ of sampling rate $F_s$ is sub-sampled from $F_s \times T_x$ to $N_x$, with $N_x<<F_s \times T_x$, and projected into a quantized space composed of \textbf{$K$} vectors ($c_i$). $K$ corresponds to the size of the musical alphabet, and the $c_i \in \mathbb{R}^{D}$ are vectors corresponding to the musical classes, i.e. tokens. For simplicity's sake, we write $\{c_1,c_2,\ldots,c_{K}\}$ = $C_K$, also called \emph{codebook}. 

\begin{equation}\label{eq:formule_perception}
\begin{aligned}
    P : \ &\vec{\textbf{x}} &\longmapsto \ &P(\vec{\textbf{x}}) = \vec{\textbf{x}}_q = [x_{q_1},x_{q_2},\ldots,x_{q_{Nx}}]\\
    &\mathbb{R}^{F_s \times T} &\longrightarrow \ &\{c_1,c_2,\ldots,c_{K}\}^{N_x}
\end{aligned}
\end{equation}

\subsection{Decision: learning and exploiting symbolic relationships}\label{sec:architecture_decision}
The musical relationships at the heart of this research are approached at a symbolic level. With the symbolic abstraction layer provided by $P$, these relationships are expressed through the relationships between two symbolic sequences extracted from two audio sequences. In the proposed formulation, this relationship between symbolic sequences is learned by the Decision module. During inference, the Decision module generates a \emph{symbolic specification} — given a conditioning input $\vec{\textbf{x}}_q$ — of what must be played by the Action module.

The symbolic specification is generated by an autoregressive model $D$ formalized by Equation \ref{eq:formule_decision}. $D$ predicts at each frame $t$ the probabilities over the $K$ tokens of the musical alphabet, given the previous token vectors $\vec{\textbf{y}}_{q<t} \in C^{t-1}_K$ in the sequence and a conditioning sequence $\vec{\textbf{x}}_q$. Then, a specific token amongst the ones with high probability can be selected for the next iteration. 
%Note that the final probability of $\vec{\textbf{y}}_{qt}$ given by Equation \ref{eq:formule_decision} is equivalent to Equation \ref{eq:coupling}.

\begin{equation}\label{eq:formule_decision}
\begin{aligned}
\centering
    D : &\vec{\textbf{x}}_q, \vec{\textbf{y}}_{q<t} & \longmapsto & D(\vec{\textbf{y}}_{q<t},\vec{\textbf{x}}_q) = p(\textbf{y}_{qt} | \vec{\textbf{y}}_{q<t},\vec{\textbf{x}}_q) \\
    & C_K^{N_x} \times C_K^{t-1} & \longrightarrow  & \mathbb{R}^{K}
\end{aligned}
\end{equation}

\subsection{Action: converting a symbolic specification to audio}\label{sec:architecture_action}
The Action module allows the agent to interact with its user and environment. This module allows switching from the symbolic domain to the audio domain: generating the audio signal $\hat{\textbf{y}}$ corresponding to the symbolic specification $\vec{\textbf{y}}_q$ given by $D$. There are many approaches to synthesizing or generating audio from a sequence of symbolic classes. However, what is expected of the Action module is to convert a sub-sampled symbolic sequence to its corresponding up-sampled audio-rate signal.

Equation \ref{eq:formule_action} describes the mathematical function realized by the Action module : given a symbolic specification $\vec{\textbf{y}}_q$ (expressed as a sequence of tokens or their corresponding class IDs), $A$ converts a sequence of discrete values back into its corresponding audio signal $\hat{\textbf{y}}$. 

\begin{equation}\label{eq:formule_action}
\begin{aligned}
    A : \  &\vec{\textbf{y}}_q &\longmapsto \ & A(\vec{\textbf{y}}_q) = \hat{\textbf{y}}  \\
    &C_K^{N_y} \text{ or } [0;K-1]^{N_y} &\longrightarrow \ &\mathbb{R}^{F_s \ast T_y}
\end{aligned}
\end{equation}

\section{Model implementation}\label{sec:model}
\begin{figure}[ht]
\centering
    \includegraphics[width=13.9cm]{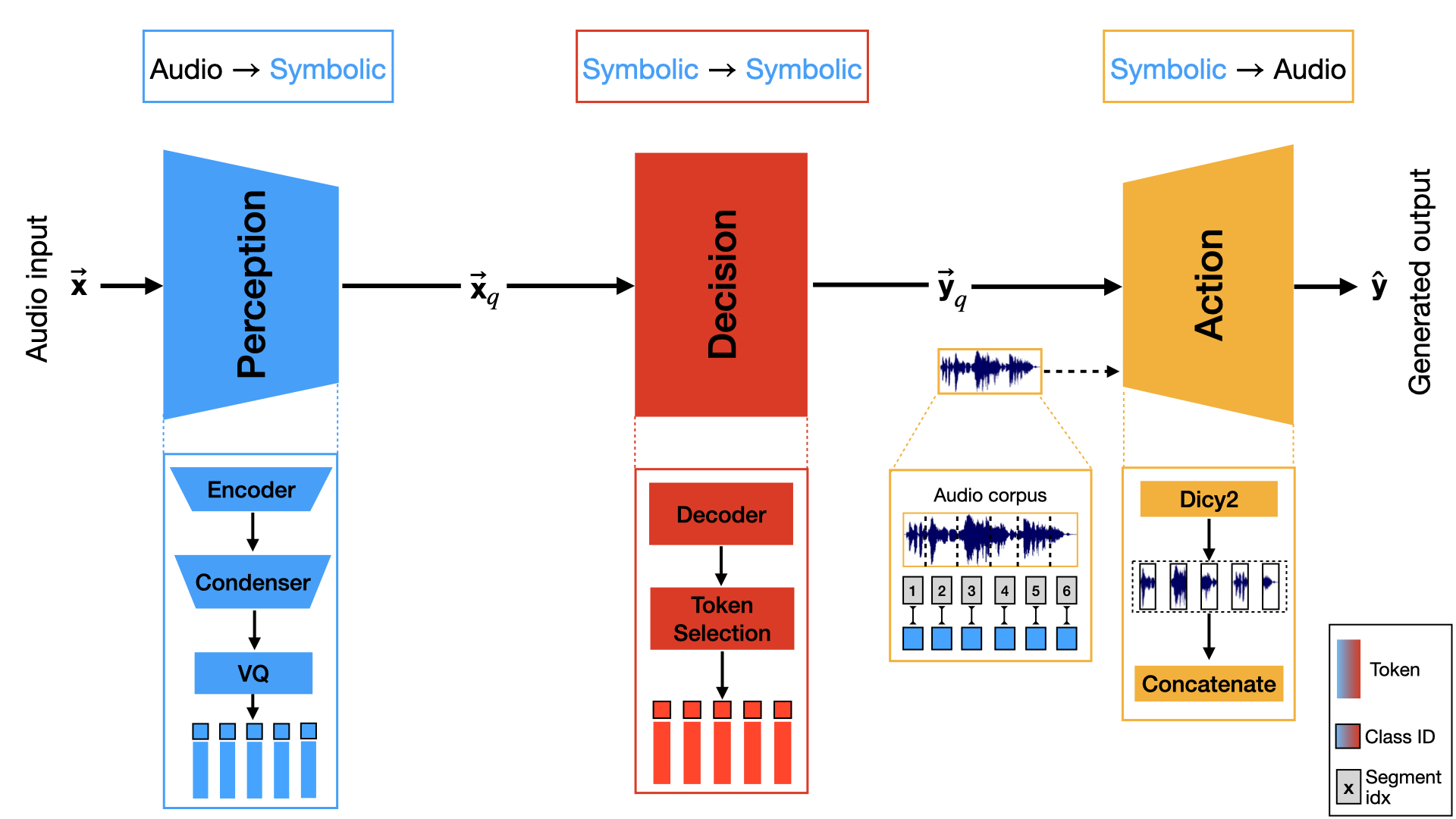}
    \caption{Overview of the proposed relationship-based Musical Agent architecture. The pipeline is divided into three modules. The Perception module encodes an audio input $\vec{\textbf{x}}$ into a sub-sampled, quantized symbolic sequence $\vec{\textbf{x}_q}$ using a pre-trained encoder, a temporal condenser, and a vector quantizer. The Decision module generates a symbolic response $\vec{\textbf{y}}_q$ based on $\vec{\textbf{x}}_q$ through a sequence decoder and token selection. Finally, the Action module reconstructs audio from $\vec{\textbf{y}}_q$ via corpus-based concatenative synthesis. The synthesis corpus is a user-defined audio file segmented and encoded through the same Perception module.}
    \label{fig:model_implementation}
\end{figure}

This section describes a specific implementation of the high-level architecture for a Musical Agent capable of generating musical responses based on relationships learned from a corpus of separate tracks\footnote{Code and audio examples can be found at \url{https://github.com/ircam-ismm/learning-from-paired-tracks}.}. Following the guidelines of Section \ref{sec:architecture}, Figure \ref{fig:model_implementation} depicts the implementation of the three components enabling the generation of such responses: a Perception module encoding audio streams into discrete musical representations on which symbolic musical relationships will be learned, an Action module converting a symbolic sequence back into audio, and finally, the Decision module which learns and exploits symbolic relationships between related tracks enabling the generation of musical responses extending beyond a matching rule between input and output. 

The overall processing pipeline consists of:
\begin{enumerate}
    \item The audio input, i.e. guide or musical context, $\Vec{\textbf{x}}$, provided as a waveform, is segmented into pieces of uniform or variable size. In the following, only uniform segmentation is studied.
    \item The Perception module encodes the whole sequence and, in the sub-sampled latent space, each segment is condensed to extract the quantified representation $\Vec{\textbf{x}}_q$.
    \item $\Vec{\textbf{x}}_q$ is processed by the Decision module to generate the symbolic specification $\Vec{\textbf{y}}_q$.
    \item The Action module uses corpus based concatenative synthesis, and exploits this specification to construct the musical response, $\hat{\textbf{y}}$.
\end{enumerate}

The internal structure of each module will be detailed in the following. 

\subsection{Perception: Wav2Vec2.0 as a music encoder}\label{sec:perception_implementation}
%Perception
The Perception module is responsible for transforming the audio signal into relevant and discrete musical representations to constitute the musical alphabet on which the relationships are learned. It consists of a self-supervised and pre-trained model (\textit{Encoder}) that extracts a contextualized representation rich in musical information, a \textit{Condenser} module to limit the computational load of the Decision module and assign a single vector to each segment, and a vector quantization (\textit{VQ}) module to assign an item of the musical alphabet to each element of the sub-sampled sequence.

%implementation
A Wav2Vec 2.0 model pre-trained on music \citep{wav2vec_music}, herein named wav2vec$_{mus}$, is chosen as the encoding model. The \textit{Condenser} consists of a simple temporal average since it does not need any training. The \textit{VQ} is implemented using a k-means algorithm trained offline for each model configuration, i.e. dataset, duration of audio segment, and vocabulary size.

%processing pipeline
The whole audio sequence $\Vec{\textbf{x}}$ is first encoded by wav2vec$_{mus}$ to extract a contextualized latent representation, then each sub-sampled segment is condensed on the time axis by the \textit{Condenser}, and finally quantized by the \textit{VQ} to generate the symbolic sequence $\Vec{\textbf{x}}_q$ and its corresponding labels; respectively \textit{Tokens} and \textit{Class IDs} in Figure \ref{fig:model_implementation}.

%\begin{figure}[ht]
 %   \centering
  %  \includegraphics[width=7.8cm]{figures/decision_implementation.jpeg}
   % \caption{Decision module.}
    %\label{fig:decision_implementation}
%\end{figure}

\subsection{Decision: Transformers generating symbolic specifications}\label{sec:decision_implementation}

%Decision
The Decision module is in charge of learning and exploiting those symbolic relations between separate musical tracks, and generating a symbolic specification. This symbolic specification is what allows for the formulation of musical responses possibly beyond the scope of matching. 

%architecture
In order to achieve this objective, the Decision module is composed of two elements: a sequence generator (\textit{Decoder}) and a token selection algorithm (\textit{Token Selection}). 

%implementation
Since the main objective of the Decision is to predict a sequence of symbols (akin to tokens), the choice of implementing it through a Transformer-decoder architecture seemed natural. The Transformer-decoder outputs a sequence of probabilities over the musical alphabet following Equation \ref{eq:formule_decision}, conditioned by $\Vec{\textbf{x}}_q$ the tokens of the encoded audio input and the previously generated tokens ($\vec{\textbf{y}}_{q<t}$). Then the selection algorithm simply selects an item from the vocabulary given its probability distribution. There are multiple choices of decoding scheme, here a top-P nucleus sampling \citep{topP} was preferred.

\subsection{Action: Dicy2 as a compositional tool for offline audio generation}\label{sec:action_implementation}

%\begin{figure}
%    \centering
%    \includegraphics[width=7.8cm]{figures/action_implementation.jpeg}
%    \caption{Action module}
%    \label{fig:action_implementatin}
%\end{figure}

%In order to leverage the symbolic specification provided by the Decision, we utilize the Dicy2 environment \citep{dicy2}. Dicy2 is a Musical Agent that utilizes symbolic \emph{scenarios} and a musical \emph{corpus} for music generation. A scenario is defined as a sequence of symbols and the corpus is defined as a sequence of (content, symbol). The scenario can be generated from the audio analysis of a musician's play for real-time improvisation, or specified during offline music composition. Dicy2 functions as a symbolic sequence generator, thereby allowing the content and symbol to assume any form. This system learns temporal relationships within its corpus and, when presented with a symbolic specification, it generates the optimal sequence of content. 
The action module of the proposed implementation is based on Dicy2 and concatenative synthesis. The action module receives as input 1) a symbolic specification resulting from the Decision module, 2) a user-defined corpus in which audio segments will be searched and rearranged to perform concatenative synthesis. 

The corpus on which the Dicy2 agent trains is extracted from an audio file processed by the Perception module, and the scenario corresponds to the output of the Decision module, i.e. the symbolic specification. The symbols and contents correspond, respectively, to the class IDs from the Perception analysis, and the segment indexes of the audio segments. 

When presented with a symbolic specification, Dicy2 generates an optimal sequence of segment indexes from the audio corpus. The musical response is then constructed by concatenating the audio segments corresponding to the slice indexes returned by Dicy2.

In the context of an action module based on concatenative synthesis, generating audio from a symbolic specification requires that the predicted labels \textit{exist} within the corpus' labels. To address this, we introduce a method called \textit{constrained generation}, which guides the decoding algorithm by favoring the selection of labels present in the corpus. Specifically, since the corpus is given as a parameter, The existing labels are available. This information can be given to the decision module and if any of those labels are present in the top-P selection, sample from them. This effectively improves the probability to generate symbols from the corpus and avoiding non-existing symbols that translate to silent audio segments with corpus-based concatenative synthesis. While it improves the portion of correct matches, it can lead to lower diversity in the symbolic specification and therefore more repetition in the musical response.
%Specifically, the set of valid labels is extracted from the corpus — given as parameter — and if any of those symbols are present in the top-P selection, sample from them. 
%\note[BB]{This effectively improves the probability to generate symbols from the corpus and avoiding non-existing symbols that translate to silent audio segments with corpus-based concatenative synthesis. While it improves the portion of correct matches, it can lead to lower diversity in the symbolic specification and therefore more repetition in the musical response.}

%\begin{figure}[ht]
 %   \centering
  %  \includegraphics[width=7.8cm]{figures/perception_implementation.jpeg}
   % \caption{Perception module.}
    %\label{fig:perception_implementation}
%\end{figure}

%One assumption that allows us to retain this method is that Transformer architectures generate contextualized sequences, and therefore if the audio information is similar then these vectors should be similar and the average of these vectors could actually reflect the global information of the segment. Finally a vector quantization (VQ) stage is used to constitute the discrete musical alphabet. Here, the proposed VQ is a k-means trained offline

\section{Methodology}\label{sec:method}

\subsection{Dataset}
%data specifications
In order to enable the learning of musical relationships, it is essential that the utilized dataset contains multi-stem mixes. The underlying argument posits that, within a mix, all the stems are inherently interconnected, thereby rendering it possible to extract musical relationships from these mixes.

%data description
Two datasets were used: MoisesDB \citep{moisesdb}, comprising 240 tracks ($\sim$60h of stems) of contemporary music with 11 instruments, and the proprietary MICA dataset, with 179 tracks ($\sim$50h of stems) of free improvisations featuring duos and trios. These contrasting music corpora allow the exploration of different musical relationships. Given the preponderance of Western pop music in MoisesDB, which is commensurate with tonal music, it can be posited that the predominant relationships in this corpus pertain to melody and harmony. Conversely, the MICA dataset is constituted of a series of non-idiomatic improvisations, which primarily encompass relationships that are predicated on interactions involving density and energy. 

To focus on narrative and discursive interaction, rhythmic elements (e.g., drums, percussion) and effects were excluded, as they do not serve the same musical function and could bias the model toward another types of relationship mainly defined by the synchronicity between musical elements.
%Two distinct dataset were employed : MoisesDB \citep{moisesdb} and the MICA proprietary dataset. MoisesDB comprises 240 audio files (around 60h all stems combined) of contemporary music, featuring 11 distinct instruments. The MICA dataset consists of 179 audio files (around 50h) derived from free jazz improvisation sessions featuring duos and trios. The utilization of these two distinct datasets will facilitate the exploration of different types of musical relationships : melodic and harmonic relationships with MoisesDB, energy and density relationships with MICA. Our  study focuses on the narrative and discursive aspects of musical interaction. For this reason, rhythmic elements, such as drums and percussion, along with effects, were excluded from the database as they did not share the same discursive role as the retained instruments and could bias the model toward synchronous elements. 

We use an 80-10-10 split for training, validation, and test sets, with audio tracks segmented into 15-second windows with 5-second overlaps. During training, a random pair of stems (A, B) is extracted from the original mix. Therefore, no unique mapping from one track to another is forced, the decision should learn symbolic relationships from a corpus-specific scale relationships. In other words we consider high-level common relationships occurring between \emph{all} pairs of the corpus. This is thus in contrast to track-specific relationships, which are relationships from a pair of input and output stem, i.e. paired instruments relationships. 

%Rhythmic and percussive elements, including drums and percussion, as well as effects, have been removed from the database. It was determined that these musical functions associated with rhythmic elements (and effects) did not share the same discursive role as the retained instruments. Furthermore, their substantial representation would significantly influence the model towards synchronous elements. Whereas this study focuses on the narrative and discursive aspect of musical interaction. We employ a 80-10-10 split for training, validation and test sets. The audio tracks are segmented with 15 seconds windows and 5 seconds of overlap. During training, a random pair of \emph{input} and \emph{ground truth} stems are extracted from the original mix. Therefore, no unique mapping from one track to another is forced, the decision has to internalize symbolic relationships from a corpus-level scale; and not track-specific relationships.

It should be noted that the size of the datasets are relatively small for the type of learning task we consider. In comparison, Diff-A-Riff \citep{diff_a_riff} trains on 1M pairs of input and target audio (equivalent to 2.7K hours), SingSong \citep{singsong} also trains its model on 1M audio inputs (equivalent to 46K hours), and MusicGen-Stem \citep{musicgen_stem} leverage 17K hours of instrumental audio. In our case, the datasets can be smaller since we do not synthesize audio from the symbolic specification. The effect of the dataset size will be discussed in Section \ref{sec:symbolic_perf_discussion}.

\subsection{Learning musical relationships}\label{sec:learning_method}

The objective of the proposed Musical Agent, and more specifically its decision module, is to learn musical relationships from a corpus of separated tracks to enable the generation of responses exploiting these learned relationships. The methodology employed in this paper to learn such relationships consists in exploiting multi-stem datasets to extract general musical relationships specific to each musical corpora. Please note that the Decision module is not designed to learn track-specific relationships.

\begin{figure}[ht]
    \centering
    \includegraphics[width=10cm]{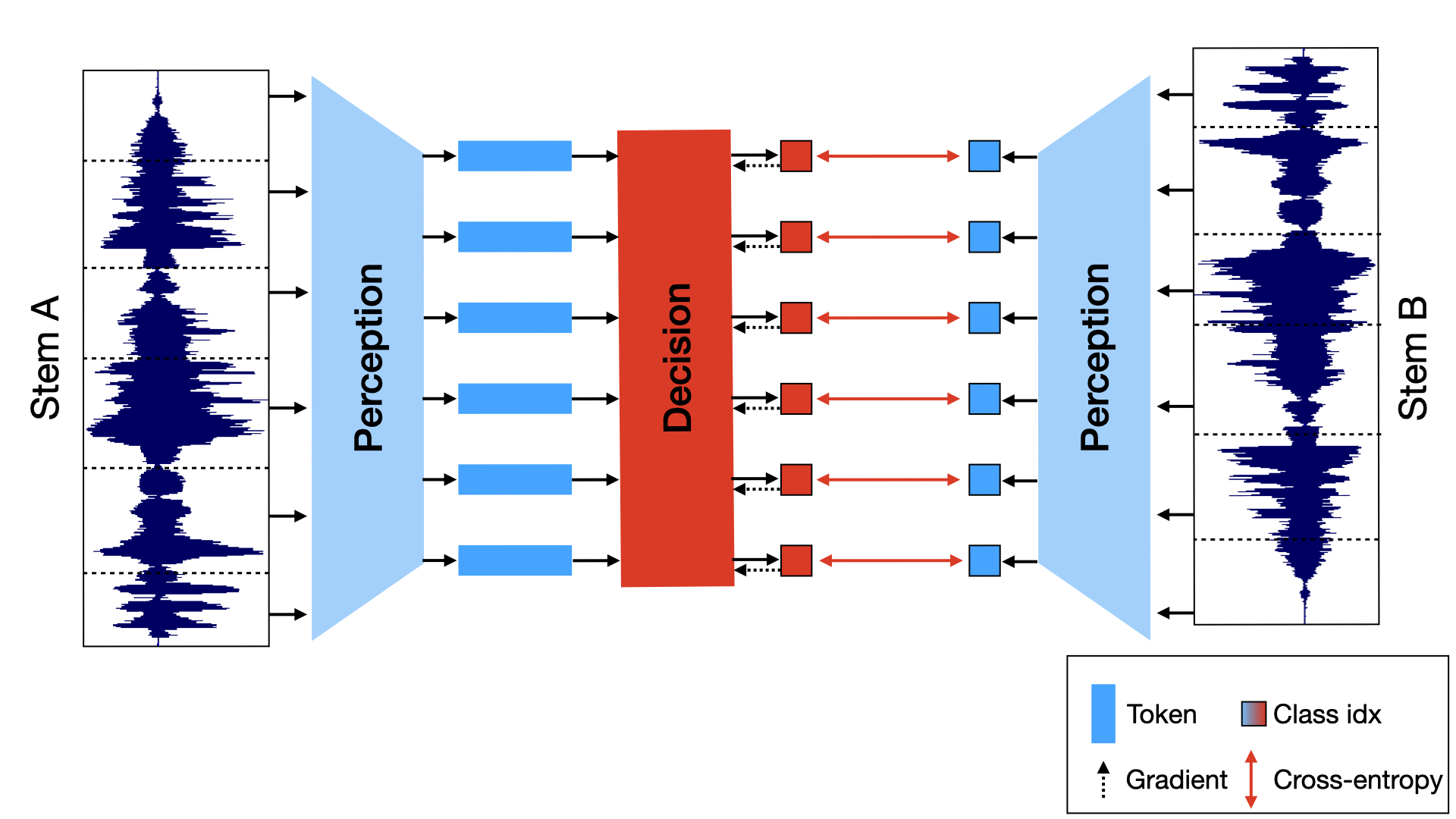}
    \caption{Training procedure for learning symbolic musical relationships.}
    \label{fig:learning_relationships}
\end{figure}

Figure \ref{fig:learning_relationships} depicts the learning process of musical relationships within 2 stems. The Perception module encodes both stems coming from the original pair (A, B), and the Decision's objective is to predict the classes associated with tokens of Stem B from the sequence of tokens of Stem A. This objective is achieved by implementing a cross-entropy loss between the predicted classes and the classes from the encoding of the original Stem B. In accordance with the principles outlined by AudioLM, we freeze the Perception module in charge of \textit{tokenizing} the audio stream \citep{audiolm}. The Decision module has 6 transformer-decoder layers, 12 heads, model dimension of 768, feedforward dimension of 2048, and 0.1 dropout. We employ Adam \citep{adam} optimizer with batch size of 24, lr = $1e^{-6}$, $\beta$ = (0.9, 0.999), %epochs $\in$ [60;160]\footnote{Training is stopped after convergence. Depending on the configuration, the convergence varied from 60 epochs (for small vocabulary size) to 160 epochs (large vocabulary size)},
weight decay = $1e^{-5}$, and trained until convergence. To reduce Exposure Bias, an exponential decay scheduled sampling training is implemented \citep{scheduled_sampling}. Preliminary experiments proved that it effectively improves the generation quality.

This study focuses exclusively on uniform segmentation of audio, a choice that is made to facilitate synchronization issues between input and generated output audios. Three sizes of segmentation windows are employed: 250 ms, 350 ms and 500 ms. This choice of uniform segmentation rather than \emph{event}-related segmentation is sub-optimal for the quality on the final audio rendering. As our primary focus in this paper concerns the proposition about learning and exploiting symbolic relationships, the limitation due to uniform segmentation will not be further evaluated and will only be investigated in future works. %Therefore,  \change[BB]{readers and listeners are invited to keep this matter in mind as they delve into the results section}{readers and listeners are invited to take this matter into consideration as they peruse the results section.}.

The proposed architecture utilizes a musical alphabet or vocabulary, the size of which constitutes a pivotal hyper-parameter. The present study investigates therefore three vocabulary sizes: 16, 64, and 256. This size exerts a substantial influence on the vocabulary's granularity, manifesting in two correlated effects. Firstly, modeling symbolic musical relationships becomes more manageable with a limited vocabulary. Conversely, diversity in relationships is diminished. 

% \begin{hypothesis}\label{hyp:vocab_size}
%     Increasing the vocabulary size will increase diversity of generated symbolic sequences, but will conversely increase the difficulty of modeling symbolic relationships between separated tracks.
% \end{hypothesis}
%justifier ce choix de tailles ?

%plus de details sur input et output : representation données, création du mix etc

\subsection{Evaluating learned musical relationships}\label{sec:method_eval}

% \begin{table}[ht]
%     \centering
%     \begin{tabular}{l|c|c}
%         \hline
%         \diagbox{\textbf{Task}}{\textbf{Domain}}& \textbf{Symbolic} & \textbf{Audio} \\
%         \hline
%         \textbf{Reconstruction} & \makecell{Accuracy \\ WER \\ \textit{Longest Prefix} \\ Diversity \\ Perplexity} & \makecell{\textit{Music Similarity} \\ APA} \\
%         \hline
%         \textbf{Creative Generation} & \makecell{Diversity \\ Perplexity} & APA \\
%         % 
        
%         \hline
%     \end{tabular}
%     \caption{Summary of evaluation metrics}
%     \label{tab:evaluation_metrics}
% \end{table}

%There is no absolute ground truth to which to compare the generated sequence in such creative task \citep{musical_agents}; whether in the symbolic or audio domain.Therefore, Musical Agents are best evaluated through qualitative and ethnological approaches. Nonetheless, as a first step, we propose to evaluate quantitatively some aspect of the generated symbols.

This section presents the evaluation methodology employed during this work. Our hypothesis is that the capacity of the decision module to exploit learned relationships can be measured through a \emph{re-generation} task. This re-generation task consists in evaluating the decision module’s ability to generate the symbolic representation of track B, conditioned on track A — (one of) the track(s) it was originally paired with in the dataset.\footnote{We remind again that this pairing of tracks (A, B) is not unique: during training, the decision learns the mapping from A to B, from B to A and all other combinations of mappings depending on the number of stems in the original mix.} This task is evaluated through two quantitative measures: \textit{True Positive Percentage} and \textit{Longest Common Prefix}.

%, such as accuracy, \textit{Word Error Rate} (WER), entropy and \textit{Longest Common Prefix}. However, we remind that the system has not been trained to extract track-specific relationships; but rather generic, corpus-specific relationships. Therefore, it is not expect from the decision engine to be able to perfectly re-generate track B from track A.

%Accuracy is the a widely used metric for classification systems (e.g. ImageNet \citep{imagenet} competition), in this context accuracy can be employed to quantify the similarity of predicted and ground truth symbolic sequences. WER is a metric used in Natural Language Processing (NLP) tasks, such as machine translation \citep{machine_translation} and automatic speech recognition \citep{asr}. It is similar to accuracy since it measures the similarity of two sequences but it can take account for unequal size sequences; accounting for discrepancies as deletion and insertion. Entropy can be used to measure diversity in a sequence, in this study we measure the entropy of the sequences in bits. Entropy can be particularly helpful to compare results across vocabulary sizes and verify Hypothesis \ref{hyp:vocab_size} of Section \ref{sec:learning_method}. 

\textit{True Positive Percentage} (TPP), expressed in Equation \ref{eq:true_positive_percentage}, is simply the count of True Positive (TP) predictions divided by the total length of the ground truth (GT) sequence. In this context, TPP is employed to quantify the similarity between the predicted symbolic sequence B', generated from stem A, and the original symbolic sequence of B from the paired stems (A, B).

\begin{equation}\label{eq:true_positive_percentage}
    \operatorname{True Positive Percentage} = \frac{\sum^{N}_{i=1} \{1 \text{ }|\text{ } pred[i] = GT[i]\}}{N}\cdot100\text{, where $N$ = length($GT$)}
\end{equation}

To complement TPP's insights on the decision's performances, we introduce the \textit{Longest Common Prefix} (LCP) measure to compute the lengths of correct prefixes between the predicted symbolic sequence B', generated from stem A, and the symbolic sequence of stem B (stem A's original pair). This metric is inspired by sequence matching in bioinformatics \citep{bio_string_algo}, in particular the Longest Common Factor \citep{lcf, lcf2} (LCF), which searches for susbequences maximizing an alignment score. LCP differs from LCF by requiring the two sequences to be aligned in time when computing the longest prefix, which is essential in this re-generation task. LCP returns a list of common prefix lengths for each position in the sequence, which provide complementary insights on the performances of the symbolic re-generation task.

Let \( S, T \in [0;K-1]^* \) be two strings of length \( n \), and let \( i \in \{0, 1, \dots, n-1\} \).  
We define the Longest Common Prefix between the suffixes \( S_i = S[i..] \) and \( T_i = T[i..] \) as:

\[
\operatorname{LCP}_i(S, T) = \max \left\{ \ell \in \mathbb{N} \,\middle|\, i + \ell \leq n,\ \forall k < \ell,\ S_i[k] = T_i[k] \right\}
\]

\section{Results}\label{sec:results}

The results of the re-generation task are presented below, measured quantitatively using True Positive Percentage (TPP) and Longest Common Prefix (LCP). Since this task has never been published with similar framework, no baseline model is available for comparison. Nevertheless, it is worth introducing a baseline, where symbolic sequences are generated by randomly sampling from the entire alphabet. Given that the task focuses on learning relationships at the corpus level rather than track-specific ones, perfect re-generation of stem B from its originally paired stem A is not expected. However, this task should demonstrate the decision's ability to at least partially learn relationships between paired tracks. 
%The random baseline serves to verify that some correspondences have been learned from the corpus. 

The $p$ parameter of nucleus sampling is set to 0.8 for all configurations. Both the model and random baseline are evaluated with and without the constrained generation (see \ref{sec:action_implementation}) for TPP measures, and only under constrained generation for LCP measures. %To assess statistical significance, we perform a Mann-Whitney U test \citep{mann_whitney}, \add[FB]{the distribution being not normal}. \note[FB]{not sure it's necessary}The  results are reported in Tables \ref{tab:canonne_p-values} and \ref{tab:moises_p-values} of the Appendix.

\begin{figure}
     \begin{subfigure}[b]{0.87\textwidth}
         \centering
         \includegraphics[width=\textwidth]{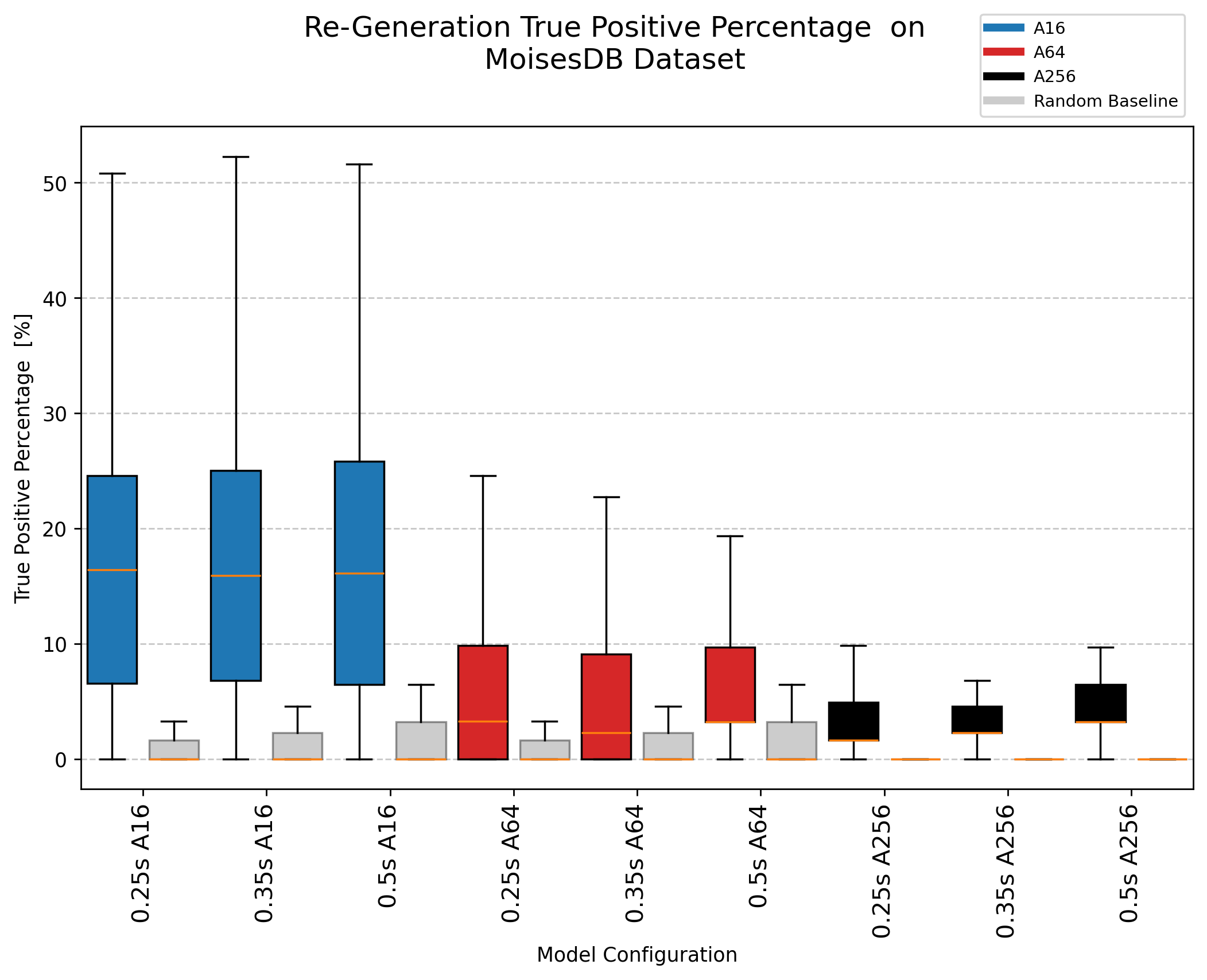}
         \caption{MoisesDB}
         \label{fig:moises_acc}
     \end{subfigure}
     \hfill
     \begin{subfigure}[b]{0.87\textwidth}
         \centering
         \includegraphics[width=\textwidth]{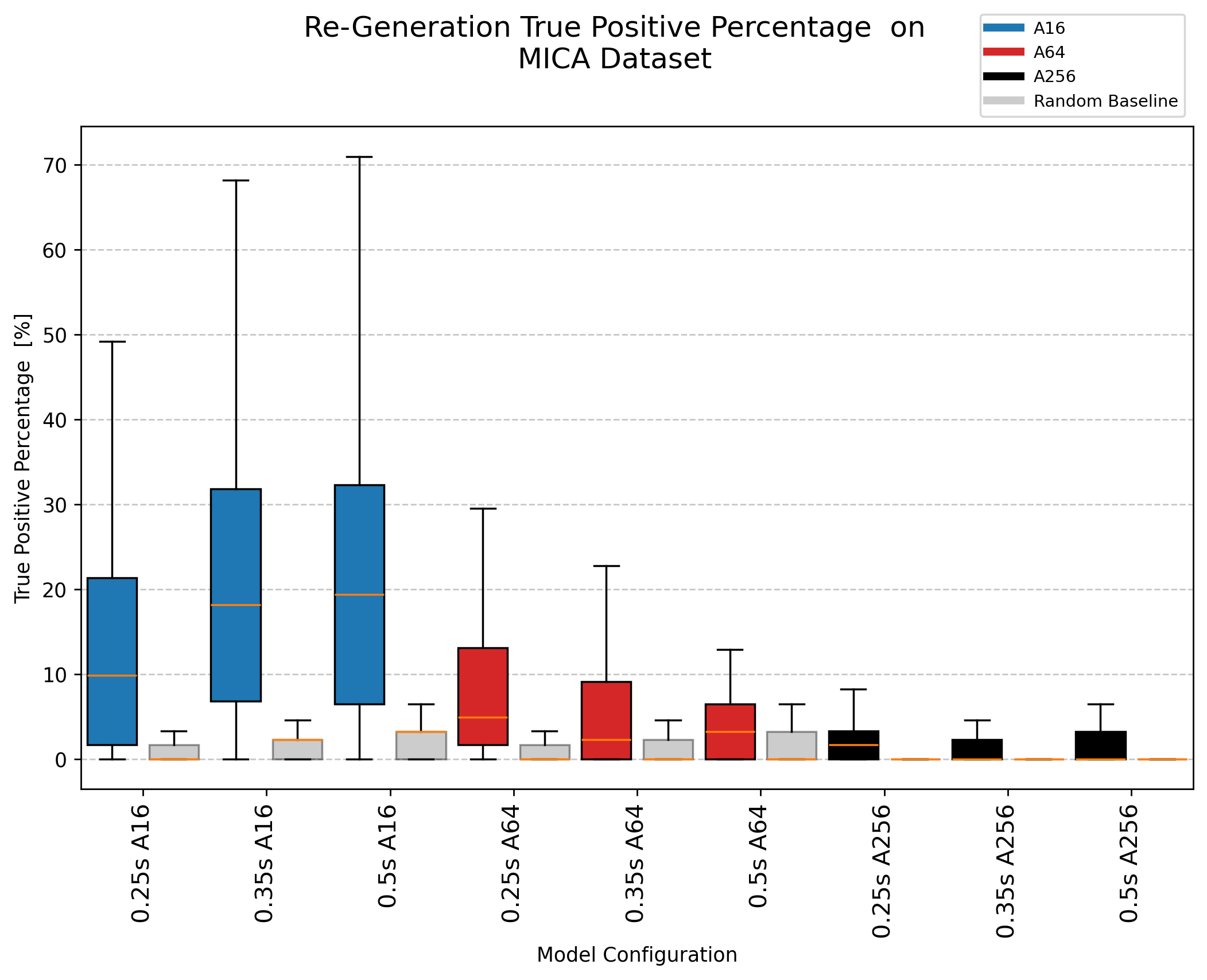}
         \caption{MICA}
         \label{fig:canonne_acc}
     \end{subfigure}
        \caption{True Positive Percentage (TPP) of the re-generation task for MoisesDB (a) and MICA (b). Model configurations are grouped by alphabet size and ordered by segmentation duration. Random baseline (gray) is plotted on the right of its corresponding configuration. Mann-Whitney statistical test returns p-value<0.0001 for all configurations.}
        \label{fig:accuracy}
\end{figure}

\begin{figure}
     \centering
     \begin{subfigure}[b]{0.87\textwidth}
         \centering
         \includegraphics[width=\textwidth]{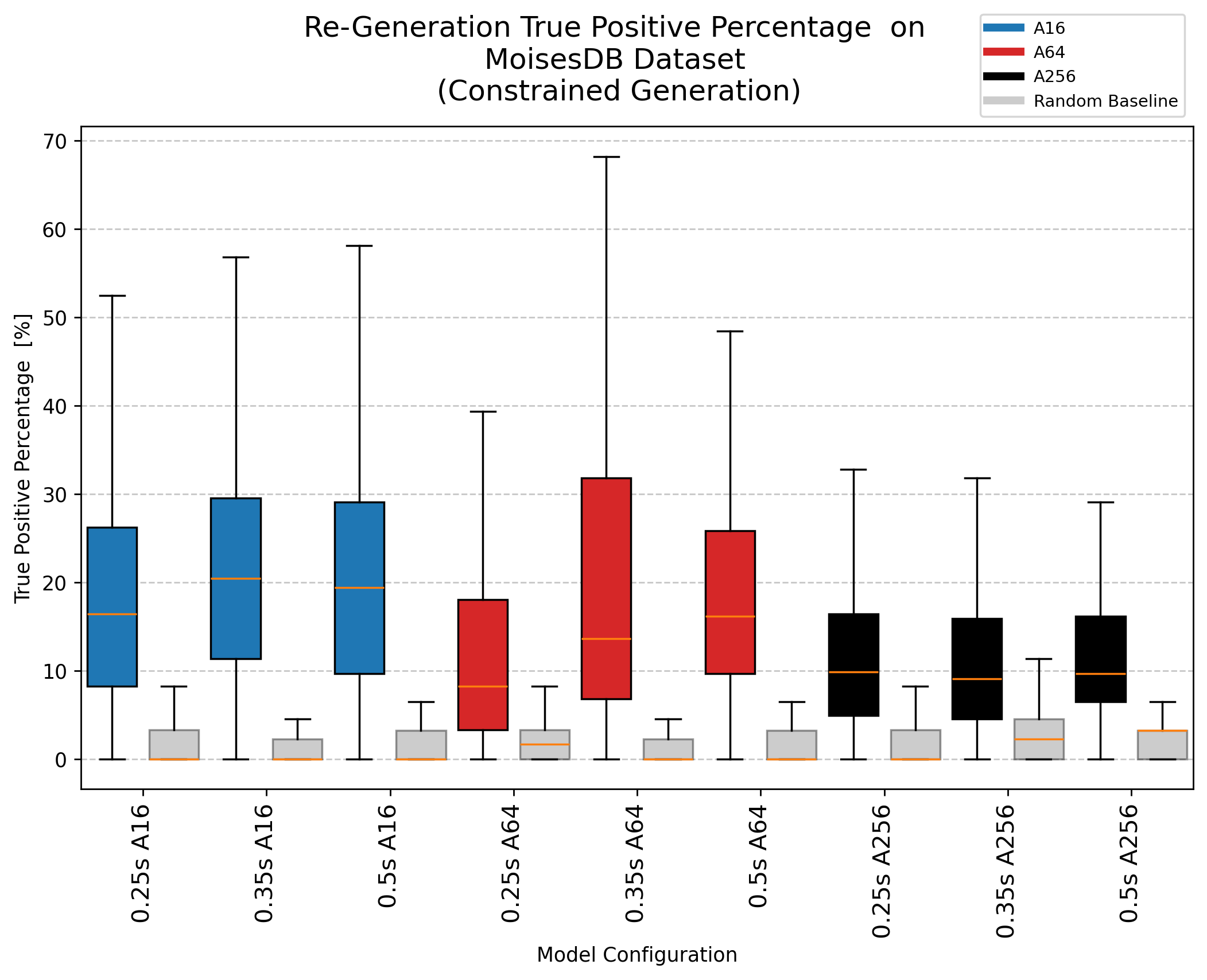}
         \caption{MoisesDB}
         \label{fig:moises_acc_fe}
     \end{subfigure}
     \hfill
     \begin{subfigure}[b]{0.87\textwidth}
         \centering
         \includegraphics[width=\textwidth]{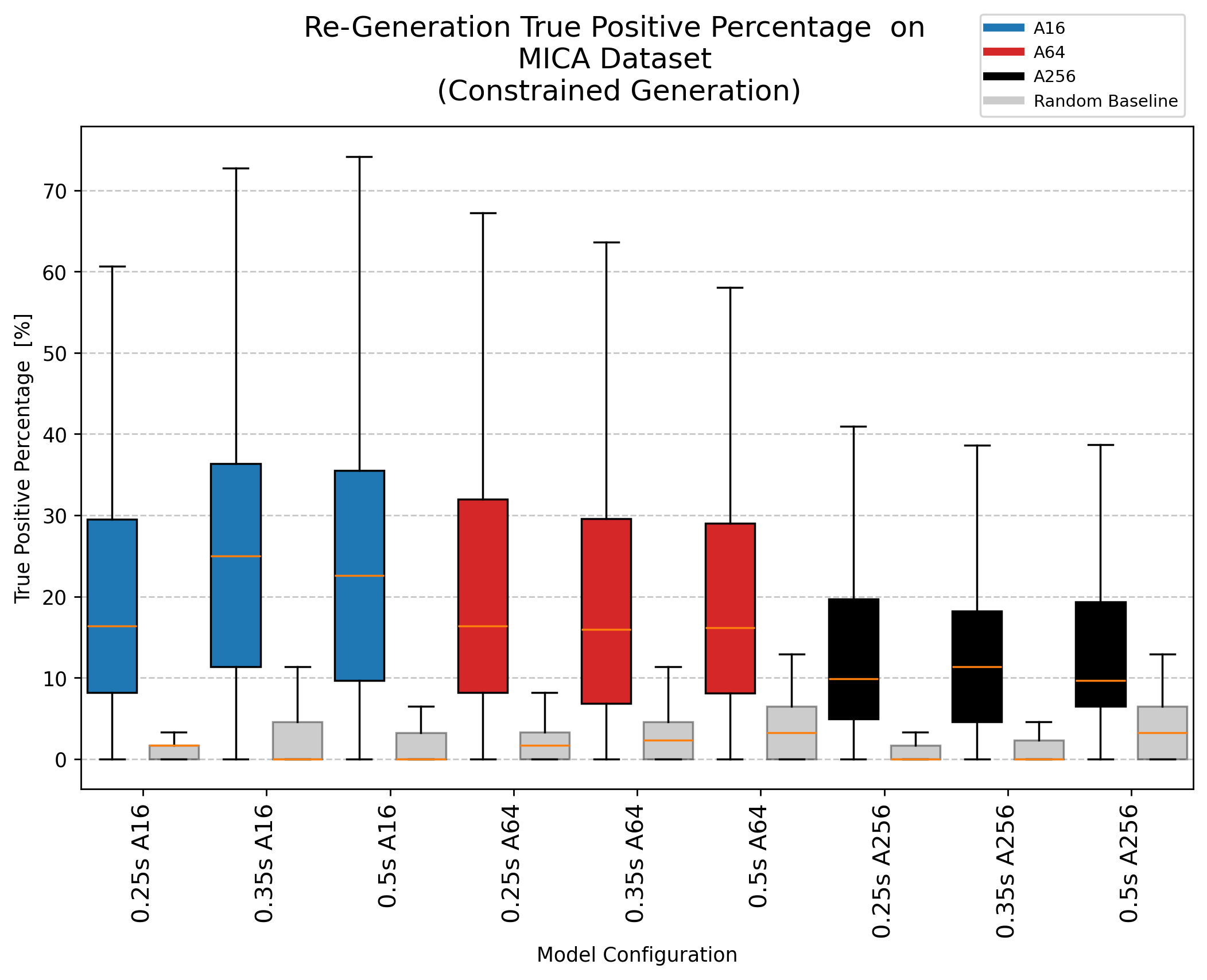}
         \caption{MICA}
         \label{fig:canonne_acc_fe}
     \end{subfigure}
    \caption{True Positive Percentage (TPP) of the re-generation task for MoisesDB (a) and MICA (b) under constrained generation setup. Model configurations are grouped by alphabet size and ordered by segmentation duration. Random baseline (gray) is plotted on the right of its corresponding configuration. Mann-Whitney statistical test returns p-value<0.0001 for all configurations.}
    \label{fig:accuracy_fe}
\end{figure}

%present figure and tell what to see : place setting for discussions
Figure \ref{fig:accuracy} presents boxplots of the TPP distributions for the MoisesDB and MICA datasets under the standard setup, while Figure \ref{fig:accuracy_fe} shows the same metric under constrained generation. Different alphabet sizes are grouped by color, with gray representing the random baseline. On the x-axis, model configurations follow the convention \{segment duration\}s\_A\{alphabet size\}. Overall, all configurations outperform random baselines. Figure \ref{fig:accuracy} shows better performances for small alphabet size, but distributions presents greater variance. While in Figure \ref{fig:accuracy_fe}, Constrained generation appears to enhance the performance of model predictions across configurations, particularly for larger alphabet sizes. For this matter, only this setup is considered for the LCP evaluation.

%Figures \ref{fig:accuracy_fe}, \ref{fig:wer_fe} and \ref{fig:diversity_fe} present respectively accuracy, WER and diversity distributions with boxoplots for ANR-MICA and MoisesDB corpora under constrained generation setup. Figures \ref{fig:accuracy}, \ref{fig:wer} and \ref{fig:diversity} correspond to the same evaluation without constrained generation. Different alphabet sizes are grouped by color, and gray corresponds to the random baseline. The x-axis corresponds to the model configuration, with the convention \{segment duration\}s\_A\{alphabet size\}. 

\begin{figure}
     \centering
     \begin{subfigure}[b]{0.94\textwidth}
         \centering
        \includegraphics[width=1\textwidth]{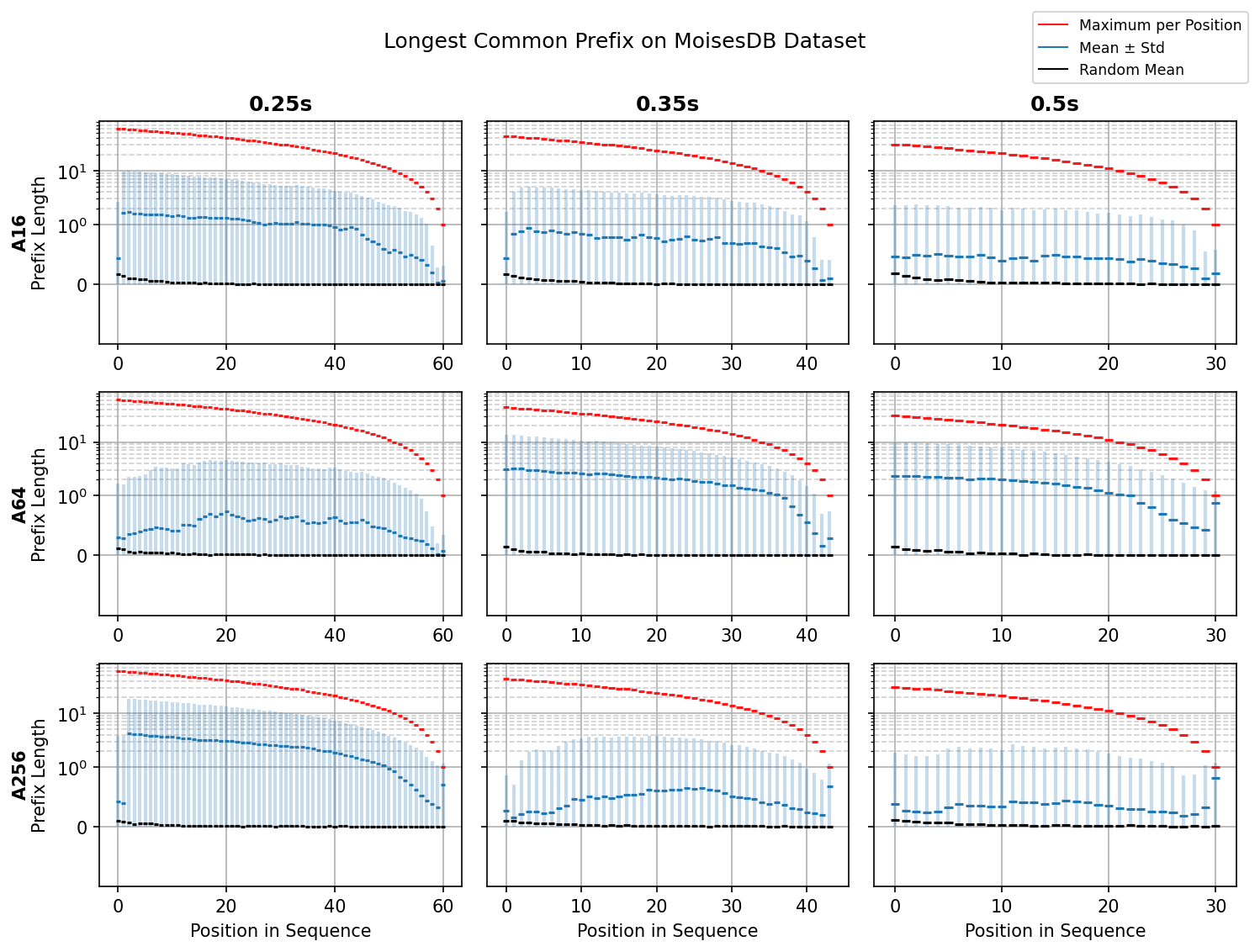}
        \caption{MoisesDB}
        \label{fig:moises_lcp_fe}
     \end{subfigure}
     \hfill
     \begin{subfigure}[b]{0.94\textwidth}
         \centering
        \includegraphics[width=1\textwidth]{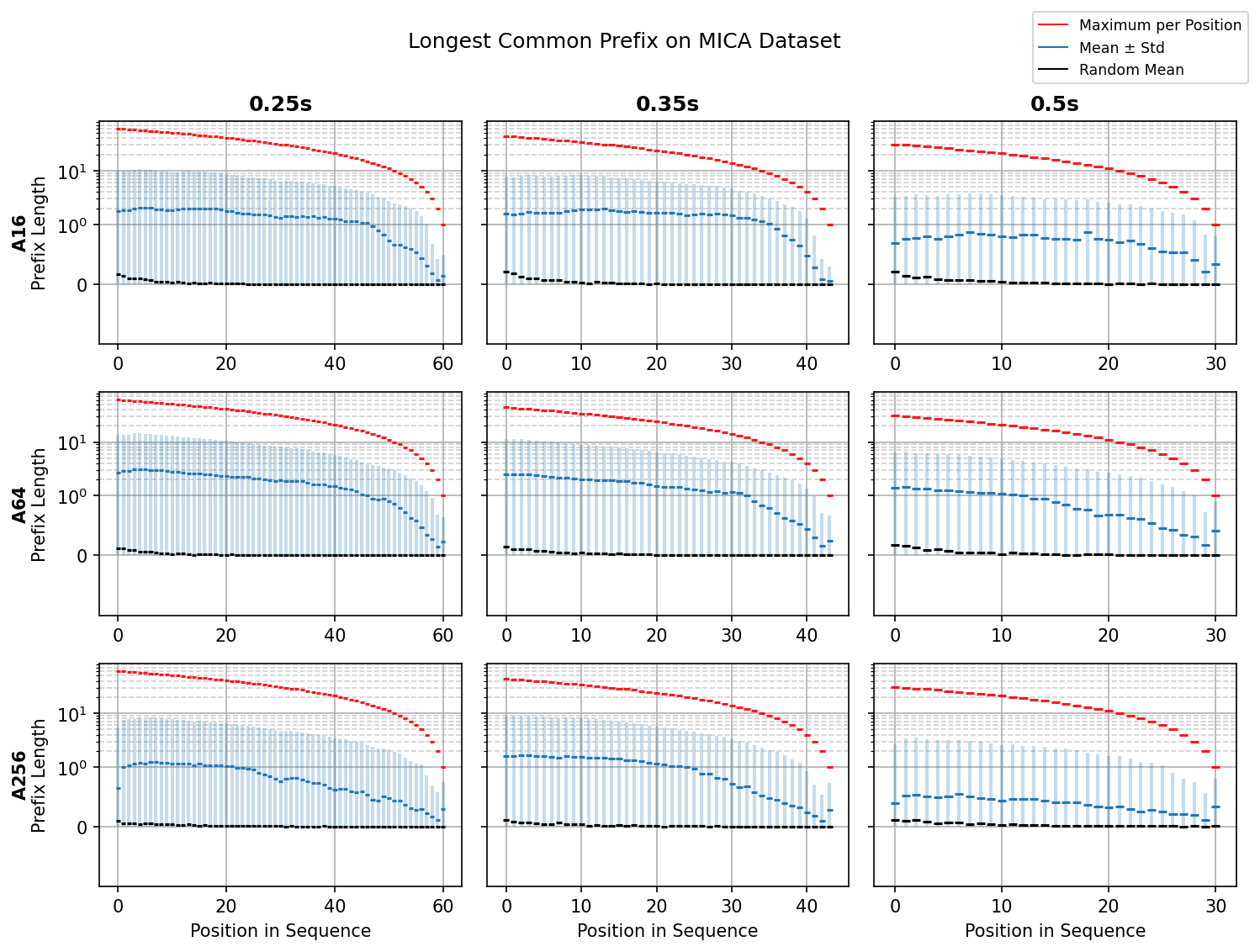}
        \caption{MICA}
        \label{fig:canonne_lcp_fe}
    \end{subfigure}
    \caption{LCP of re-generation task for MoisesDB (a) and MICA (b) under constrained generation setup. Model configurations are ordered by alphabet size (vertical axis) and segment duration (horizontal axis). Blue lines represent the mean prefix length, red lines are the maximum prefix length, and black lines represent the mean of the random baselines.}
    \label{fig:lcp}
\end{figure}

Figures \ref{fig:moises_lcp_fe} and \ref{fig:canonne_lcp_fe} show the LCP values per sequence position for, respectively, MoisesDB and MICA datasets under the constrained generation setting. Both figures present the nine configurations, ordered by alphabet (y-axis) and segmentation size (x-axis). The blue lines represent the mean LCP with standard deviation in light blue, while the black lines show the random baseline for comparison. The red line indicates the maximum possible LCP at each position, calculated as $L-x$, where $L$ is the sequence length and $x$ the position. A symmetric logarithmic scale is used to better visualize low LCP values and highlight small variations across the sequence. In general, LCPs tend to exceed the random baseline, yet they continue to demonstrate difficulty in approaching the maximum value.

%Figures \ref{fig:canonne_lcp_fe} and \ref{fig:moises_lcp_fe} are the LCP plots for, respectively, ANR-MICA and MoisesDB under constrained generation. Figures \ref{fig:canonne_lcp} and \ref{fig:moises_lcp} are the corresponding plots without constrained generation. These figures present the mean (bar plot) and standard deviation (whiskers) of the LCP at each position in the sequence. The choice to shift right each timestep helps to visualize the actual coverage of the LCP at each position in the sequence; rather than having all LCP(s) on the same x coordinate and only visualize local coverage.

%---No FC----
%---FC----

% \begin{figure}
%     \centering
%     \includegraphics[width=1\textwidth]{LaTeX/figures/evaluation/moises_lcp_fe_2.png}
%     \caption{LCP of re-generation task under constrained generation setup for MoisesDB dataset}
%     \label{fig:moises_lcp_fe}
% \end{figure}

% \begin{figure}
%     \centering
%     \includegraphics[width=1\textwidth]{LaTeX/figures/evaluation/canonne_lcp_fe_2.png}
%     \caption{LCP of re-generation task under constrained generation setup for MICA dataset}
%     \label{fig:canonne_lcp_fe}
% \end{figure}

\section{Discussion}\label{sec:discussion}
This section offers key interpretations of the results presented above. We begin with synthesized analysis of our results on the re-generation task in the beginning of \ref{sec:re_generation_discussion}, prior to a per-metric analysis in Sections \ref{sec:tpp_eval} and \ref{sec:lcp_eval}, which sheds light on several aspects of the re-generation task. This is followed by a broader discussion in Section \ref{sec:symbolic_perf_discussion}, identifying core limitations in our current architecture and methodology, and outlining directions for future work on relationship-based responses in Musical Agents.

\subsection{Re-generation task}\label{sec:re_generation_discussion}

This work explored the possibility to include learned relationships in the decision process of Musical Agents. The training of the different model configurations converged for both datasets, and the symbolic performances on the re-generation task prove that some relationships have been learned. The evaluation on each dataset leads to \textit{True Positive Percentage} (TPP) between 10\% and 20\% of the learned relationships. Such percentage values, which might appear as low, was expected, since the TPP values do not take into account any specificity of the different tracks, but rather to general "musical relationships" that are emerging globally in each database. Thus, the important finding is that we can confirm such general relationship are learned, as shown by the fact that the values are significantly larger than a random generation (p-value < 0.0001). Therefore, what has been learned must be high-level, common relationships among all paired tracks, i.e. corpus-level relationships. Moreover, such learning appears in both stylistically very different datasets, MoisesDB and MICA, related to different music genre (pop/rock and free improvisation). %but there is a considerable inner-corpus diversity regarding musical relationships. 
%Therefore, what has been learned must be high-level, common relationships among all paired tracks, i.e. corpus-level relationships.
This observation allows us to, at least partially, validate the interest of our framework. In the following sections we discuss in more details how the quantitative metrics True Positive Percentage and Longest Common Prefix are influenced by different parameters.

\subsubsection{True Positive Percentage}\label{sec:tpp_eval}

Results show that smaller vocabularies significantly outperform random baselines, while larger vocabularies lead to a noticeable performance drop. The segmentation size appears to have little impact for the TPP. Instead, increased vocabulary size generally correlates with reduced performance, likely due to the difficulty of modeling fine-grained relationships. %Particularly for MICA dataset with median values close to 0 for A256.
Interestingly, MoisesDB maintains more stable results even with larger alphabets, suggesting more coherent or consistent relationships within its corpus. Constrained Generation significantly improves the TPP for larger vocabularies by prioritizing symbols already present in the paired sequence to which it is compared to. On average, this subset represents only $6\pm2\%$\footnote{Measured on the validation set} of the vocabulary in the A256 configuration, effectively restricting predictions to a small portion of the total vocabulary and thereby increasing the likelihood of correct matches. However, the large variance observed across configurations (as shown by the wide boxplots) indicates a pronounced variability in the model's performances. 

To complement our results, we also report an entropy measure of the generated sequences to quantify the diversity of the sequences in Figures \ref{fig:diversity} and \ref{fig:diversity_fe} of the Annexes. Generally, diversity increases with alphabet size, an expected result since a larger alphabet implies larger maximum entropy.
%To complement our results, we also report audio evaluation metrics (Audio Prompt Adherence \citep{apa} and Music Similarity \citep{music_similarity}) in Table \ref{tab:audio_eval} of the Appendix, which further confirm that generated responses are more similar to ground truth than random outputs. %Nevertheless, they are paradoxically less coherent in context—raising questions about the interpretability and relevance of these metrics.

\subsubsection{Longest Common Prefix}\label{sec:lcp_eval}

Models which are closer to the red baseline (max) indicate better results and LCP analysis shows that our results outperform random baselines, which is a good indicator of the success of our approach. These results remain difficult to be fully explained, nevertheless we propose below some interpretations.

First, while results outperform random baseline, the prefix lengths remain low, indicating difficulty in anticipating future tokens. 
%and reliably predicting the current timestep. 
%LCP values are typically higher at the start of sequences and decline over time, as expected due to the shrinking maximum prefix length. The final token (EOS) slightly breaks this trend but still proves difficult to predict despite its fixed position.

We observe that the smallest segmentation window performs best for the MICA dataset, while no similar conclusion can be drawn for MoisesDB. In particular, for MICA, A64/0.25s yields the highest LCP, whereas A256/0.5s performs the worst. This suggests that small segmentation windows are better at capturing fine-grained information, likely because MICA is composed of free improvisations where information variation depends on individual notes.

In contrast, MoisesDB—composed of Western pop music, i.e., \textit{pulsed} music—appears to require a segmentation window that aligns with the beat (pulse) of the track. While LCP shows no consistent trend across segment duration or alphabet size for MoisesDB, we find that A256/0.25s approaches the maximum LCP value, whereas A256/0.5s is only slightly better than the random baseline.

Across both datasets, we note that a large vocabulary combined with a large segmentation window performs poorly. This may indicate that encoding long audio segments with high class granularity generates incoherent or unreliable information that is more difficult to model. These findings suggest that encoded information in small segments is more reliable and easier to model, especially in datasets with fine temporal variation.

Unlike TPP, LCP shows no consistent trend across segment duration or alphabet size. Notably, strong TPP does not guarantee high LCP: the 0.25s A256 configuration yields the best and most stable LCP scores for MoisesDB, whereas the 0.5s A16 — strong in TPP — performs poorly in LCP. Although not consistent across segment durations, A256 tends to offer competitive LCP results, possibly due to lower variance in TPP and the constrained prediction space improving consistency.

Consistent LCP would suggest a model’s capacity to anticipate, a key trait for real-time musical applications where forward-looking behavior ensures coherent interaction \citep{these_nika}. Further investigation is required to assess this consistency : particularly, further analysis of LCP distribution is required in order to characterize since the variance in our results does not enable to draw clear conclusions.

\subsection{Limitations}\label{sec:symbolic_perf_discussion}

%musical relationships consist of one-to-many mappings
%While we can show that some symbolic relationship are learned, we could globally expect much better results. Several factors in the training process may help explain these results. 
First, the nature of musical relationships itself is inherently complex, thus refined learning and evaluation strategies are necessary. In particular, %, as it involves a one-to-many relationship, which makes it more challenging to achieve consistent and accurate outputs.
the metrics used for evaluation are quite strict and reflect only partially the validity of our architecture. These metrics compare the generated output to a single, fixed ground truth, which may not capture the diversity of valid outputs that could be considered "perceptively" correct. Moreover, this also limits the assessment of the model's flexibility and its ability to generalize. Importantly, it would be necessary to evaluate qualitatively the performance of the system.

The second limitation is the combination of a small dataset with high intrinsic variability. 
%Music is a complex, multi-layered signal, encompassing diverse levels of abstraction \citep{music_tonal_description}, melodic patterns, harmonic or rhythmic structures, instrument, levels of expressivity \citep{instrument_classification}, timbre \citep{caracterisation-du-timbre, perception-musical-timbre}, and genre. Even within our stylistically constrained corpora, this variability makes it difficult for the model to learn stable, consistent patterns. One can argue that larger dataset could 'mechanically' improve the learning of such relationships, the access to reliable datasets remains a significant challenge. 
While the accompaniment systems discussed in \ref{sec:gen_ai_in_music} benefit from large proprietary datasets, publicly available resources — especially multitrack datasets — remain scarce. MoisesDB \citep{moisesdb} is currently the largest multitrack dataset for pop/rock, but includes little jazz. Recent efforts such as the Jazz Trio Dataset by \citet{jazz_trio_ds}, which provides around 45 hours of multi-instrument recordings, are valuable contributions, yet the majority of available data remains heavily biased toward Western pop music \citep{ethical_biases_2,ethical_biases}.

Furthermore, the Perception module produces symbolic sequences with a class imbalance — persisting across different alphabet sizes, datasets, and segmentation durations — and sequences exhibiting long repetitions of the same token. This pattern encourage the model to overfit this repetition pattern and produce highly repetitive outputs (especially under greedy sampling strategies like $k=1$).
%a class imbalance—persisting across different alphabet sizes, datasets, and segmentation durations—and symbolic sequences exhibiting long repetitions of the same token, encourage the model to overfit this repetition pattern and produce highly repetitive outputs (especially under greedy sampling strategies like $k=1$).

% \note[BB]{Paragraphe superflu ?}
% Compounding these challenges is the nature of the learning process itself, which is driven by relationships drawn from the full corpus rather than from specific track pairs. Expecting the model to internalize one-to-one relationships between individual tracks without any form of pairwise adaptation is, in hindsight, overly optimistic. It is more plausible that the model captures higher-level, abstract patterns that generalize across the corpus, rather than fine-grained correspondences unique to each pair. 

Finally, the choice of uniform segmentation may also hinder the model's performance. Incorrectly aligned segmentation windows can result in segments that lack coherence. For instance, one half of a segment might contain silence, while the other half may have musical notes. This misalignment can create issues in terms of both coherence and semantic meaning within the generated sequences. Furthermore, the presence of long notes may span across multiple tokens, emphasizing this repetition pattern in the output of the Perception module.

One of the central challenges encountered in this study lies in the complexity of the implemented architecture, which makes it difficult to interpret the results with clarity. While the Wav2Vec2.0-based Perception module may capture musical information, the nature of this information remains opaque. Without a clearer understanding of the representations encoded at this stage, it becomes particularly difficult to evaluate what kind of symbolic relationships, if any, are effectively learned by the Transformer-based Decision module. This lack of transparency hinders our ability to determine whether the model captures meaningful musical correspondences between paired tracks.

\section{Conclusion and Future Works}\label{sec:conclusion}

We presented a general architecture for audio generation guided by the listening of another audio stream. We established the baseline for a learning model inferring symbolic relationships between a perception module and the output of a decision module.

To evaluate this baseline, we presented quantitative results on a re-generation task: generating a sequence using a decision module trained on a large dataset of paired tracks (A, B). Precisely, we compared the generated output to a reference track B, originally paired with A in the dataset. To move forward toward an architecture that is highly customizable via a curated user dataset, we will extend the quantitative analysis to a re-construction task aiming at reconstructing B strictly after fine-tuning of the decision module on the (A, B) pair.

Furthermore, our work will focus on what we could call \textit{creative generalization} — namely, attempting to identify relationships comparable to those found in the training dataset between entirely different inputs and outputs. Such questions being highly subjective, they will be evaluated using  qualitative methods in the context of fieldwork with musicians. % However, a part of the quantitative analysis will seek not to evaluate but to characterize the musical dimensions effectively learned by our decision model. 
This question is closely tied to the characterization of the musical dimensions encoded in the perception module we have developed, such as the similarity or proximity between classes established by our encoder. These aspects will also be examined in order to refine our loss functions by introducing notions of "prediction error quality", considering functional equivalence classes of tokens \citep{carsault}.

\section*{Acknowledgments}
We would like to express our gratitude to Dr. Alessandro Ragano for providing the weights of the Wav2Vec 2.0 model that was pre-trained on music.

\section*{Ethical statement}
This research did not involve any human participants or animals. The datasets used in this work were accessed legally, either under open access (MoisesDB) or formal agreement of the data copyright owners (MICA). %Thus, there are no data related issues such as copyright, privacy, or consent. 
We focused on minimizing the environmental footprint of neural network computations. The code developed for this project will be released as open-source software to ensure reproducibility. The authors report no conflicts of interest. 

\bibliographystyle{apalike}   
\bibliography{biblio}  

\appendix
\newpage

\newpage
\section{Re-generation task : diversity}

\begin{figure}[ht]
     \centering
     \begin{subfigure}[b]{0.45\textwidth}
         \centering
         \includegraphics[width=\textwidth]{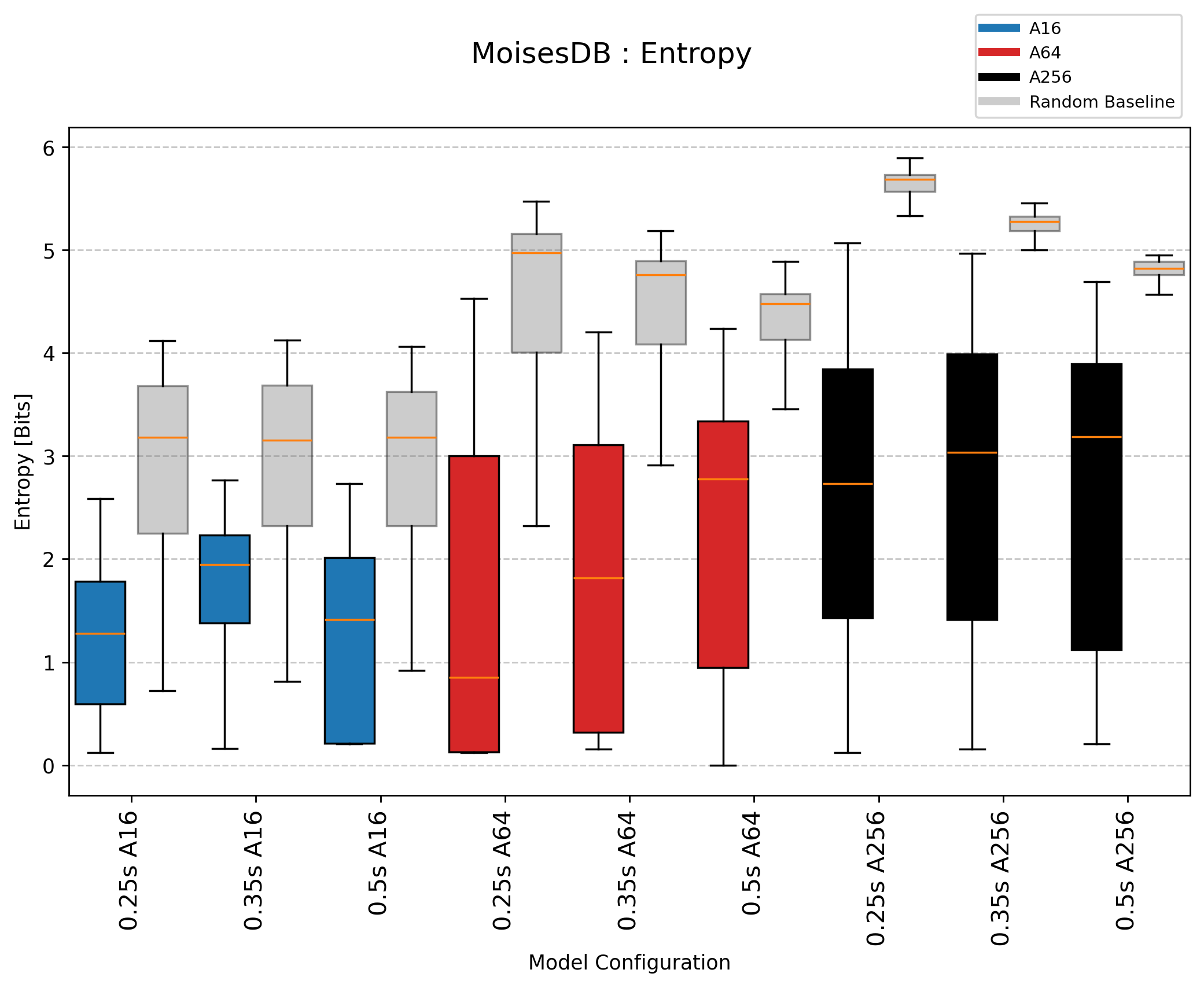}
         \caption{MoisesDB}
         \label{fig:moises_div}
     \end{subfigure}
     \hfill
     \begin{subfigure}[b]{0.45\textwidth}
         \centering
         \includegraphics[width=\textwidth]{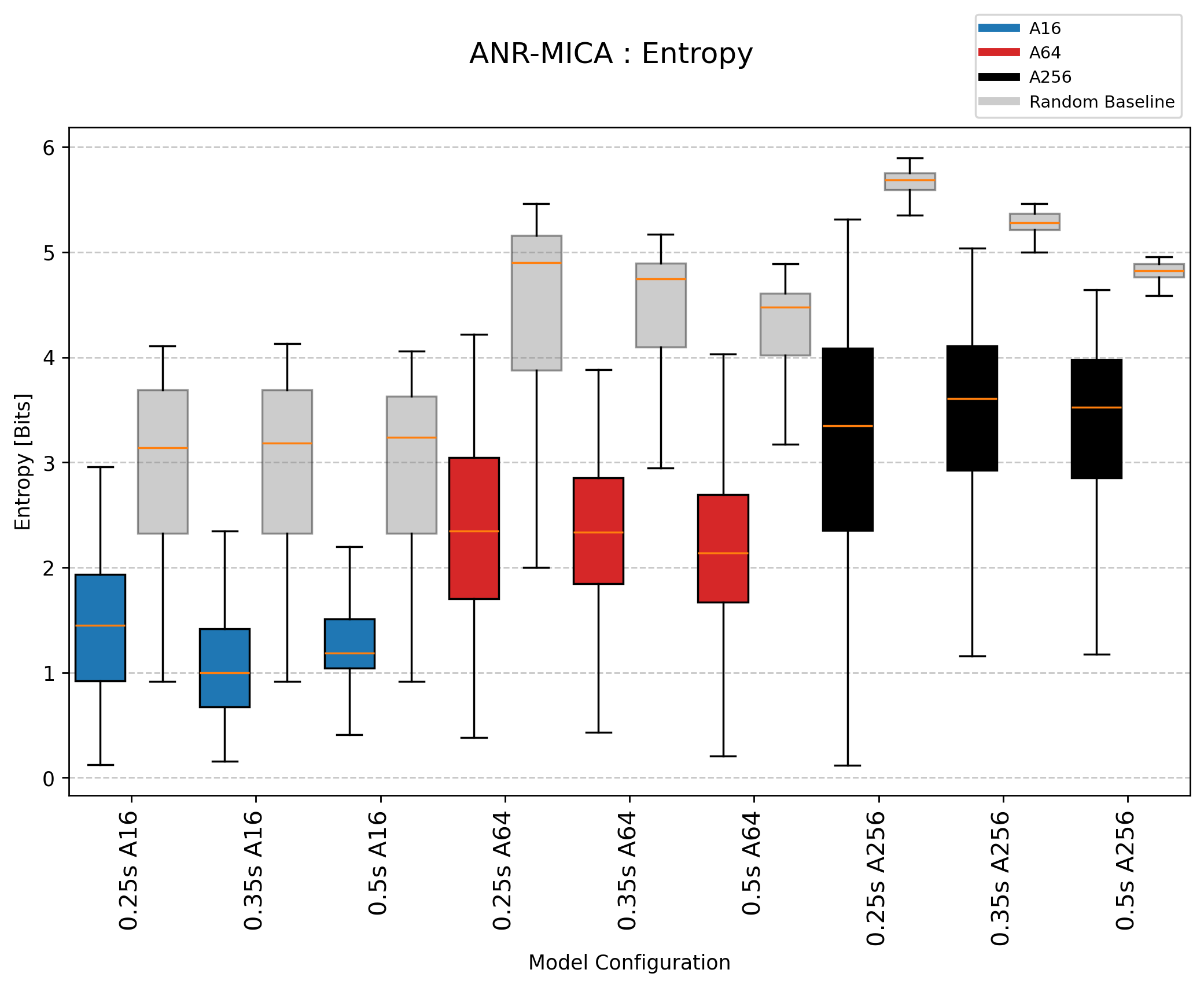}
         \caption{MICA}
         \label{fig:canonne_div}
     \end{subfigure}
    \caption{Entropy of re-generation task for MoisesDB and MICA}
    \label{fig:diversity}
\end{figure}

\begin{figure}[ht]
     \centering
     \begin{subfigure}[b]{0.45\textwidth}
         \centering
         \includegraphics[width=\textwidth]{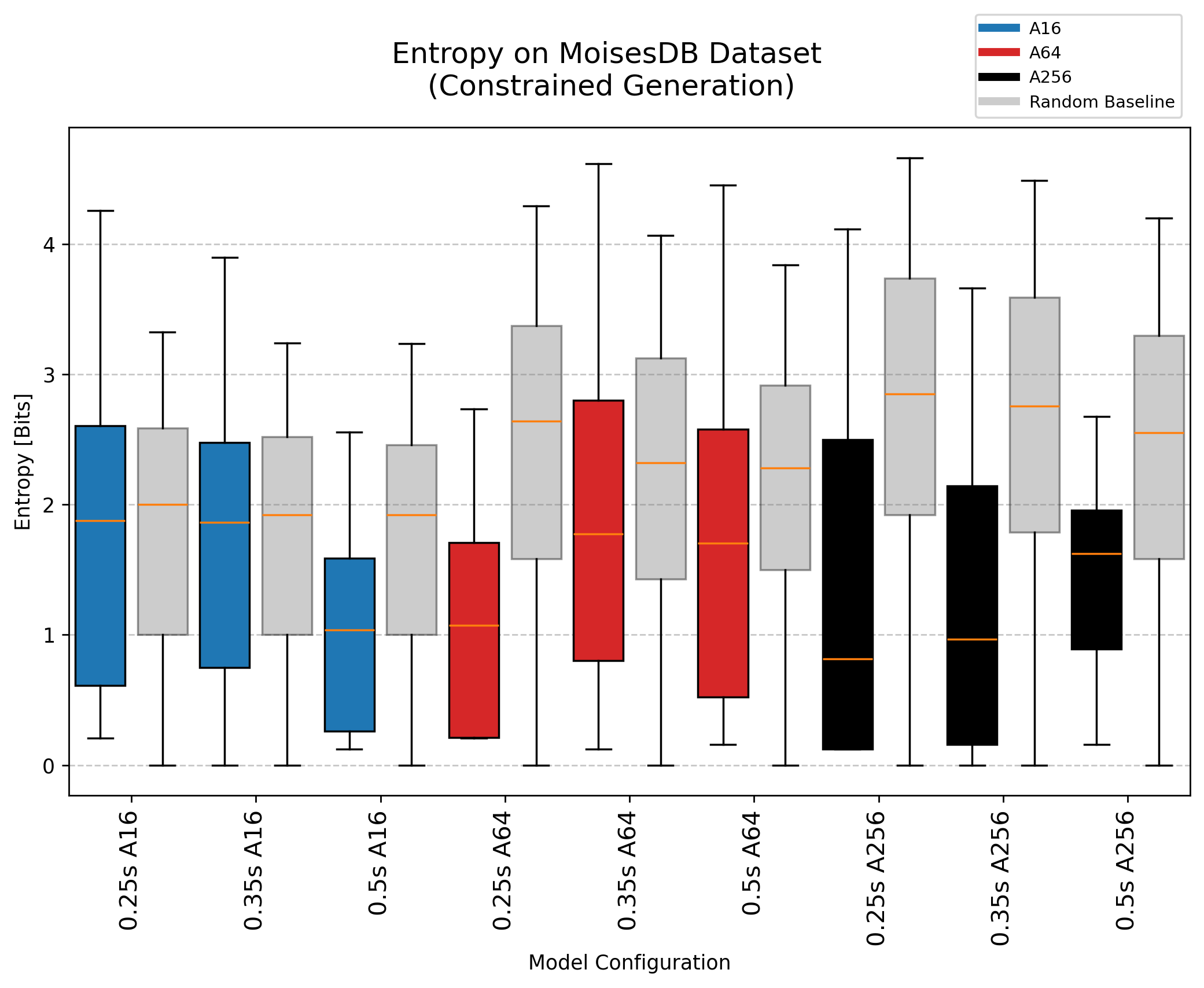}
         \caption{MoisesDB}
         \label{fig:moises_div_fe}
     \end{subfigure}
     \hfill
     \begin{subfigure}[b]{0.45\textwidth}
         \centering
         \includegraphics[width=\textwidth]{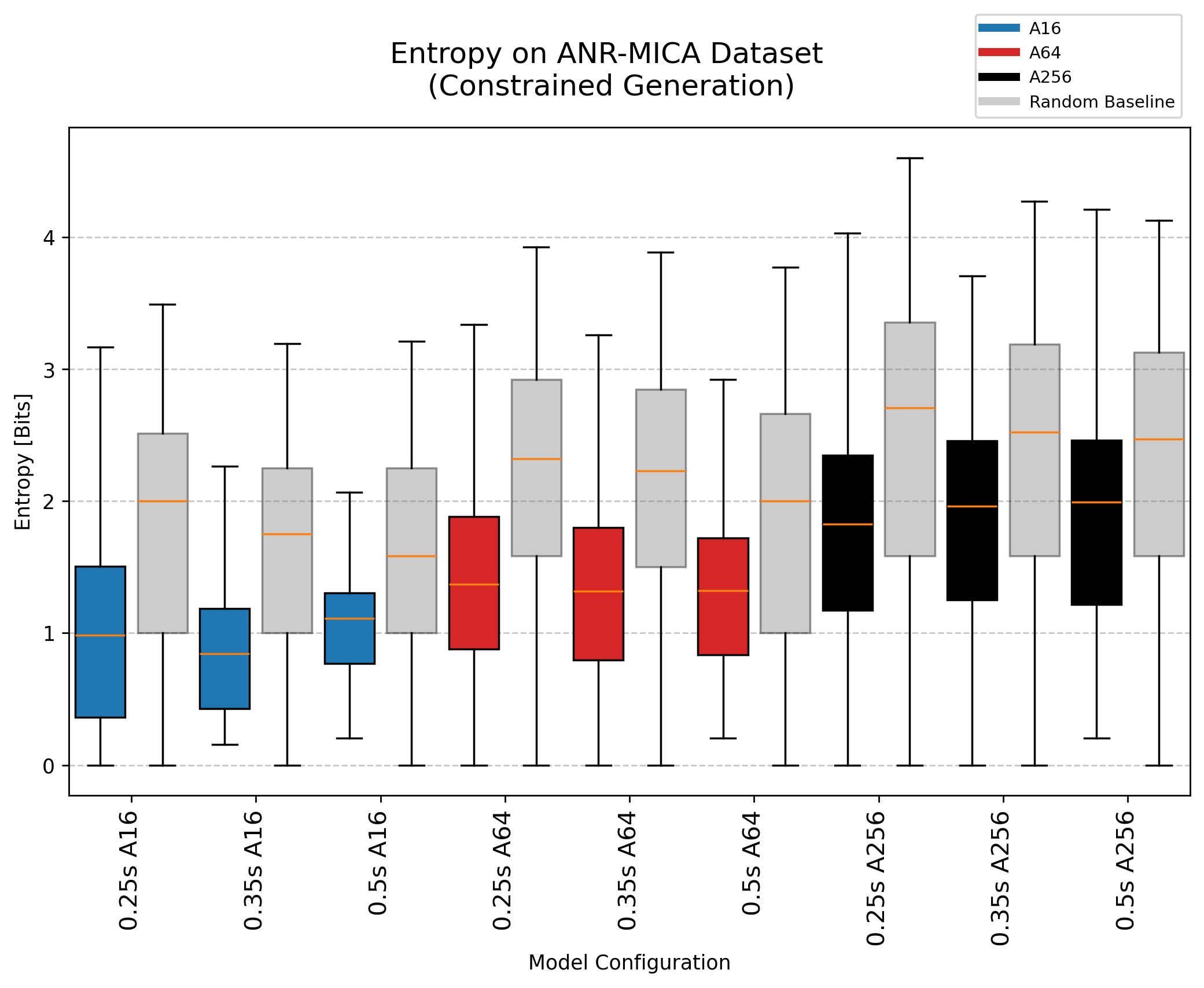}
         \caption{MICA}
         \label{fig:canonne_div_fe}
     \end{subfigure}
    \caption{Entropy of re-generation task for MoisesDB and MICA under constrained generation setup}
    \label{fig:diversity_fe}
\end{figure}

\end{document}